\documentstyle[epsf]{elsart}

\begin{document}
\def\bq{\begin{equation}}
\def\eq{\end{equation}}
\def\bqa{\begin{eqnarray}}
\def\eqa{\end{eqnarray}}
\def\roughly#1{\mathrel{\raise.3ex
\hbox{$#1$\kern-.75em\lower1ex\hbox{$\sim$}}}}
\def\lsim{\roughly<}
\def\gsim{\roughly>}
\def\llgm{\left\lgroup\matrix}
\def\rrgm{\right\rgroup}
\def\gslash#1{\slash\hspace*{-0.20cm}#1}
\renewcommand{\theequation}{\arabic{section}.\arabic{equation}}

\begin{frontmatter}

\title{GRACE/SUSY \\
 Automatic Generation  of Tree Amplitudes  in the Minimal
Supersymmetric  Standard Model}

\author[KEK]{J.~Fujimoto}
\author[KEK]{T.~Ishikawa}
\author[TMC]{M.~Jimbo}
\author[KEK]{T.~Kaneko}
\author[KGU]{K.~Kato}
\author[KEK]{S.~Kawabata}
\author[SKU]{K.~Kon}
\author[MGU]{M.~Kuroda}
\author[KEK]{Y.~Kurihara}
\author[KEK]{Y.~Shimizu}
\author[RKU]{H.~Tanaka}
\address[KEK]{KEK, Oho, Tsukuba, Ibaraki 305-0801 Japan}
\address[TMC]{Tokyo Management College, Ichikawa, Chiba 272-0001, Japan}
\address[KGU]{Kogakuin University, Nishi-Shinjuku, Tokyo 163-8677, Japan}
\address[SKU]{Seikei University, Musashino, Tokyo 180-8633, Japan}
\address[MGU]{Meijigakuin University, Totsuka, Yokohama 244-8539, Japan}
\address[RKU]{Rikkyo University, Nishi-Ikebukuro, Tokyo 171-8501, Japan}

\begin{abstract}
 {\tt GRACE/SUSY} is a program package for generating the tree-level
amplitude and evaluating the corresponding cross section of
processes of the minimal supersymmetric extension of the standard
model (MSSM). The Higgs potential adopted in the system, however,
is assumed to have a more general form indicated by the
two-Higgs-doublet model. This system is an extension of {\tt
GRACE} for the standard model(SM) of the electroweak and strong
interactions. For a given MSSM process the Feynman graphs and
amplitudes at tree-level are automatically created. The
Monte-Carlo phase space integration by means of {\tt BASES} gives
the total and differential cross sections. When combined with {\tt
SPRING}, an event generator, the program package provides us with
the simulation of the SUSY particle productions.
\end{abstract}

\end{frontmatter}

\section{Introduction}
     It is widely expected that the supersymmetric (SUSY) extension
 of the standard model(SM) of particle physics \cite{susy}
is one of the most promising approaches that could resolve the
problems remaining in the standard model.
In particular, the minimal
supersymmetric extension of the standard model (MSSM) has been extensively
studied in the last decade, because it has the simplest structure
containing the least number of particles,
yet it is complex enough to allow the most essential features
characteristic in any theory of SUSY.

Over these ten years we  have been working with the automatic
computation of  high energy physics processes.  The system for the
computation of the SM, {\tt GRACE}, has been published in
\cite{grace}.  In order to be able to handle  SUSY processes, we
have made several necessary modifications and extension of the
system \cite{kon1,tkkjk}. First we have installed the new
interaction vertices according to the MSSM lagrangian allowing an
extension to a more general effective Higgs potential (section
2.4). Second we have modified the program so that we can deal with
the fermion number non-conserving vertices and Majorana particles,
both of which are inherent to any SUSY theory.

Using this extended version of {\tt GRACE/SUSY}, we have published
a package of event-generator, {\tt SUSY23}, which contains 23 specific SUSY
processes for $e^+e^-\to$ 2-body and 3-body \cite{susy23}.

Since then we have made extensive checks of the system. In this
paper we present the detailed description of the program system
which enables us to compute automatically the cross section of any
MSSM tree process, and explain the structure of {\tt GRACE/SUSY}
together with  instructions for how to use it.

Program pacakge {\tt GRACE/SUSY} (\texttt{GRACE v2.2.0}) is available
form \\ \texttt{http://minami-home.kek.jp/}.

\section{Theoretical framework of the MSSM}
\subsection{Basic lagrangian}
  The basic lagrangian of the MSSM in its full form has been given in several
papers\cite{rosiek,hikasa,mk1}.  In this work we use the convention
adopted in ref.\cite{mk1}, where the positively (negatively)
charged charginos are considered as fermion (anti-fermion).
The full form of the MSSM lagrangian is too lengthy to be reproduced here.
Readers should refer to the original work \cite{mk1}.
We show only the soft SUSY-breaking terms in order to define the SUSY
breaking parameters.
\bqa
{\cal L}_{soft} &=& -\mu \int d^2\theta \Phi_{\bf H_1}\Phi_{\bf H_2}
                     + h.c. \cr
     & & -{1\over 2}M_1\lambda\lambda
                        -{1\over 2}M_2\lambda^a\lambda^a
                        -{1\over 2}M_3\tilde g^\alpha\tilde g^\alpha + h.c.\cr
     & & -\tilde m_1^2{\bf H}_1^*{\bf H}_1
                       -\tilde m_2^2{\bf H}_2^*{\bf H}_2
                       -(\tilde m_{12}^2{\bf H}_1\llgm{ 0 & 1 \cr -1 & 0}\rrgm
                          {\bf H}_2 + h.c.) \cr
     & &- {{\sqrt 2 m_u}\over v_2}A_u {\bf H}_2{\bf A}(q_L)A(u_R)
      + {{\sqrt 2 m_d}\over v_1}A_d {\bf H}_1{\bf A}(q_L)A(d_R)+h.c.\cr
     & &+ {{\sqrt 2 m_e}\over v_1}A_e {\bf H}_1{\bf A}(\ell_L)A(e_R)
       + h.c \cr
     & & - \sum _f[ \tilde m^2_{\tilde f_L} \tilde f^*_L \tilde f_L
                  + \tilde m^2_{\tilde f_R} \tilde f^*_R \tilde f_R ].  \label{2.1}
\eqa
In the first line $\Phi_{{\bf H}_1}$ and $\Phi_{{\bf H}_2}$ stand for
the iso-doublet Higgs
superfields with hypercharge $Y=-1$ and $Y=+1$ respectively, and
in the second line $\lambda$, $\lambda^a$ and $\tilde g^\alpha$ are
the gaugino fields for $U(1)$, $SU(2)$ and $SU(3)$ gauge symmetries,
respectively.  ${\bf H}_i$ appearing in the third to fifth line are
the Higgs doublets which are decomposed as
\bqa
  {\bf H}_1\equiv\llgm{ H^0_1\cr H^-_1}\rrgm&=&
       \llgm{(v_1+\phi_1^0-i\chi_1^0)/\sqrt 2 \cr
             -\phi_1^-}\rrgm,\cr
  {\bf H}_2\equiv\llgm{ H^+_2\cr H^0_2}\rrgm&=&
       \llgm{\phi_2^+ \cr
          (v_2+\phi_2^0 + i \chi_2^0)/\sqrt 2}\rrgm.  \label{2.2}
\eqa
The scalar fields ${\bf A}(q_L)$, $A(u_R)$, {\it etc}. represent the
sfermion fields, the superpartners of the corresponding fermion
fields $\llgm{u_L \cr d_L}\rrgm$ and $\bar u_R$, respectively,
\bq
    {\bf A}(q_L)=\llgm{\tilde u_L \cr \tilde d_L}\rrgm,~~~
             A(u_R) = \tilde u^*_R,  ~~~   A(d_R) = \tilde d^*_R.
    \label{2.3}
\eq
Although the so-called $R$-parity violating interactions \cite{rparity}
are not included in this lagrangian and hence in {\tt GRACE/SUSY},
users can implement them in the system by coding the necessary
vertex functions {\it etc}. in appropriate subprograms.


\subsection{Particle content}
    The MSSM consists of superfields containing the known fermions or bosons
as their component fields, and two Higgs superfields. Note that
in contrast to the SM, the MSSM includes three neutral Higgs particles
and a pair of charged Higgs particles. The mass eigenstates of
the CP-even neutral Higgs states are defined by
\bq
   \llgm{H^0 \cr h^0}\rrgm =
   \llgm{\cos\alpha & \sin\alpha \cr -\sin\alpha & \cos\alpha}\rrgm
   \llgm{\phi_1^0 \cr \phi_2^0}\rrgm,  \label{2.4}
\eq where the rotation angle $\alpha$ is determined in such a way
that the Higgs mass matrix becomes diagonal.  Since any of the
known particles has its superpartner, the number of the elementary
particles in the MSSM is more than double the number of particles
appearing in the SM. In fact, for each known fermion $f$ (except
for neutrinos), there are two kinds of sfermions, the left-handed
sfermion $\tilde f_L$ and the right-handed sfermion $\tilde f_R$.
The mass eigenstates of sfermion $\tilde f_i$ with mass $m_{\tilde
f_i}$ ( $m_{\tilde f_1}< m_{\tilde f_2}$) is defined by the
orthogonal matrix \bq
    \llgm{\tilde f_1 \cr \tilde f_2} \rrgm =
    \llgm{\cos\theta_f & \sin\theta_f \cr -\sin\theta_f & \cos\theta_f}\rrgm
    \llgm{\tilde f_L \cr \tilde f_R}\rrgm,~~~~f=e,u,d,\cdots \label{2.5}
\eq
where the rotation angle is determined by the mass diagonalization condition.

For each gauge boson there is a gaugino, and to each Higgs particle
corresponds a higgsino.  Two neutral gauginos, $\lambda$ and $\lambda^0$,
and two neutral higgsinos, $\tilde H_1^0$ and $\tilde H_2^0$, are combined
to form four Majorana fermions called neutralinos, $\tilde\chi_i^0\equiv
\llgm{ \bar\lambda_i \cr \lambda_i}\rrgm$.
The two-component spinors $\lambda_i (i=1\sim 4)$ are  linear
combinations of $\lambda$, $\lambda^0$, $\tilde H^0_1$ and $\tilde H^0_2$,
\bq~~
    \tilde\lambda_i = \eta_i\bigl({\cal O}_N)_{ij}\chi_j,~~~~i=1 \sim 4
    \label{2.6}
\eq
where $\chi_j = (\lambda, \lambda^0, \tilde H_1^0, \tilde H_2^0)$,
and masses are ordered as $m_{\tilde \chi^0_1} \le m_{\tilde \chi^0_2} \le
m_{\tilde \chi^0_3} \le m_{\tilde \chi^0_4}$. The sign factor $\eta_i$
takes the value of $+1$ or $+i$, and ${\cal O}_N$ is the orthogonal matrix
which diagonalizes the mass matrix of the
$(\lambda, \lambda^0, \tilde H_1^0, \tilde H_2^0)$ system.

The other neutral gauginos
of the strong interaction, gluinos, remain unmixed with other particles.

Corresponding to the charged Higgs particles $H^\pm$ and charged
gauge bosons $W^\pm$, there are two charged higgsinos,
$\tilde H^-_1$ and $\tilde H_2^+$, and a pair of  charged gauginos (wino),
$\lambda^\pm$.  Two charged fermions called charginos,
$\tilde\chi^+_i= \llgm{\overline{\lambda^-_{iR}} \cr \lambda^+_{iL}}\rrgm$,
are linear combination of these particles,
\bqa
   \llgm{\lambda^-_{1R} \cr \lambda^-_{2R}}\rrgm &=&
   \llgm{\cos \phi_R & \sin\phi_R \cr -\sin\phi_R & \cos\phi_R}\rrgm
   \llgm{\lambda^- \cr \tilde H^-_1}\rrgm, \cr
   \llgm{\lambda^+_{1L} \cr \lambda^+_{2L}}\rrgm &=&
   \llgm{1 & 0 \cr 0 & \epsilon_L}\rrgm
   \llgm{\cos \phi_L & \sin\phi_L \cr -\sin\phi_L & \cos\phi_L}\rrgm
   \llgm{\lambda^+ \cr \tilde H^+_2}\rrgm. \label{2.7}
\eqa
The mixing angle $\phi_R$, $\phi_L$ and the sign factor $\epsilon_L$
are determined so that their masses satisfy the ordering
$0 \le m_{\tilde\chi^\pm_1} \le m_{\tilde\chi^\pm_2}$
\footnote{Note that the value of the mixing angles $\phi_L$ and
$\phi_R$ become unstable when $\tan\beta= -{\mu/{M_2}}$ or
$\cot\beta= -{\mu/{M_2}}$.  If this happens,
$\tan\phi_L$ or $\tan\phi_R$ become $\pm \infty$.}.

Together with the particles appearing in the SM, these SUSY-particles
constitute the ingredient of the MSSM.  The list of the particles
and their coding names is  given in Appendix A.


\subsection{Basic parameters of the MSSM}
    The basic parameters of the MSSM come in when the
lagrangian is expressed in terms of superfields \cite{mk1}.  They are


\begin{tabular}{|l|l|}\hline
   coupling constants & $g, g^\prime, g_s, \mu$ \\
   gaugino mass parameters   & $M_1, M_2,M_3$ \\
   fermion masses (for each generation) & $m_e, m_u, m_d$ \\
   Higgs mass parameters     & $\tilde m_1, \tilde m_2, \tilde m_{12}$ \\
   trilinear couplings(for each generation)     & $A_e, A_u, A_d$ \\
   sfermion mass terms(for each generation)    & $\tilde m_{\tilde e_L},
   \tilde m_{\tilde e_R},
\tilde m_{\tilde u_L},\tilde m_{\tilde u_R},
\tilde m_{\tilde d_R}$ \\ \hline
\end{tabular}


The vacuum expectation values $v_1$ and $v_2$ are determined
at the minimum of the Higgs potential and they are expressed
in terms of these basic parameters.
Assuming massless neutrinos, there are in all  4 coupling constants,
$3N_G$ fermion masses, $6+8N_G$ soft supersymmetry breaking parameters,
where $N_G$ stands for the number of fermion generations.


\subsection{The MSSM parameters in {\tt GRACE/SUSY}}
  As alternatives to ten MSSM parameters for the coupling constants and gaugino
and Higgs masses shown above, we can use in {\tt GRACE/SUSY} the following set
of ten parameters as inputs,
\bq
  \alpha_e, \alpha_s, M_W, M_Z, \tan\beta, M_{A^0}, \mu, M_1,M_2,M_3.  \label{2.8}
\eq
Unless specified explicitly by the user (see Appendix B), $M_1$ and $M_3$ are
computed from $M_2$ by the following unification conditions
\bqa
  M_1 &=& {5\over 3}\tan^2\theta_W M_2,   \label{2.9}   \\
  M_3 &=& {{g_s^2}\over{e^2}}\sin^2\theta_W M_2.  \label{2.10}
\eqa
Concerning the eight sfermion mass parameters in each
generation, we use three mixing angles, and two up-type squark masses,
the lighter down-type squark mass, and two down-type slepton
masses as independent input parameters. Due to the $SU(2)_L$
invariance of the ${\cal L}_{\rm soft}$, the heavier down-type
squark mass and the sneutrino mass are given as the solution of
the following two equations (shown here only for the third
generation sfermions),
\bqa
   \sin^2\theta_b m^2_{\tilde b_2} &=&\cos^2\theta_t m^2_{\tilde t_1}
     + \sin^2\theta_t m^2_{\tilde t_2}-\cos^2\theta_b m^2_{\tilde b_1}
     +m_b^2 -m_t^2 -M_W^2 \cos 2\beta, \cr
    & & ~~~~~~~~~~~~~~~~~~~~~~~~~~~~~~~~~~~~~~~~~~~~~~~~~~\label{2.11}\\
   m^2_{\tilde \nu_{\tau}}
   &=&~ \cos^2\theta_\tau m^2_{\tilde \tau_1}
     +\sin^2\theta_\tau m^2_{\tilde \tau_2} -m_\tau^2+ M_W^2 \cos2\beta.
   \label{2.12}
\eqa
    Since we restrict our consideration only to
the CP conserving processes, all the parameters are real.  In particular,
all the physical masses used as the input are real and positive.
The allowed range of the SUSY parameters used in {\tt GRACE/SUSY}
whose values are not yet fixed by experiments are as follows\footnote{
 In ref.\cite{hikasa} the value of $\mu$ is restricted in the range
 $0\le \mu$,  while $-\infty <\tan\beta < +\infty$ is allowed.
Due to the invariance of the lagrangian with respect to an interchange
$(\mu,\tan\beta)\leftrightarrow (-\mu,-\tan\beta)$,
the physics with $\mu=\mu_0>0$ and $\tan\beta=-\tan\beta_0<0$
in ref.\cite{hikasa} corresponds to the one with $\mu=-\mu_0<0$
and $\tan\beta=\tan\beta_0>0$ in {\tt GRACE/SUSY}.}
\bqa
   &&-\infty <\mu < +\infty,~~~0\le \tan\beta,~~~
    0\le M_1, M_2, M_3, M_{A^0},~~~ \cr
   &&0\le m_{\tilde f_1}\le m_{\tilde f_2},~~~
   -{\pi\over 2}<\theta_f \le +{\pi\over 2}.  \label{2.13}
\eqa
The CP-even Higgs, charginos and neutralinos are numbered as
\bq
   M_{h^0}<M_{H^0},~~~m_{\tilde \chi^\pm_1}<m_{\tilde\chi^\pm_2},~~~
   m_{\tilde \chi^0_1}<m_{\tilde\chi^0_2}<m_{\tilde \chi^0_3}<m_{\tilde\chi^0_4}.
   \label{2.14}
\eq

The basic parameters of the MSSM are expressed in terms of
these {\tt GRACE/SUSY}
variables given in (\ref{2.8}) as follows,
\bqa
   g &=& e \sqrt{{{M_Z^2}\over{M_Z^2-M_W^2}}}, \nonumber \\
   g^\prime &=& e \sqrt{{{M_Z^2}\over{M_W^2}}}, \nonumber \\
    A_f &=& \sin\theta_f\cos\theta_f{{m^2_{\tilde f_2}
            -m^2_{\tilde f_1}}\over {m_f}}
              -\cases{ \mu\cot\beta, & $f=u,c,t$ \cr
                       \mu\tan\beta, & $f=d,s,b,e,\mu,\tau$},\cr
    \tilde m^2_{\tilde f_L}
       &=&\cos^2\theta_fm^2_{\tilde f_1}+\sin^2\theta_fm^2_{\tilde f_2}
     -m_f^2-M_Z^2\cos2\beta(T_{3f}-Q_fs^2_W),\cr
         & & ~~~~~~~~~~~~~~~~~~~~~~~  f=\nu_e,\nu_\mu,\nu_\tau,e,\mu,\tau,u,d,c,s,t,b, \cr
    \tilde m^2_{\tilde f_R}
       &=&\sin^2\theta_fm^2_{\tilde f_1}+\cos^2\theta_fm^2_{\tilde f_2}
          -m_f^2-M_Z^2\cos2\beta Q_fs^2_W,~\cr
         & & ~~~~~~~~~~~~~~~~~~~~~~f=e,\mu,\tau,u,d,c,s,t,b.
 \label{2.15}
\eqa
In the last two equations of (\ref{2.15})  the identities
$\tilde m^2_{\tilde e_L}=\tilde m^2_{\tilde \nu_e}$,
$\tilde  m^2_{\tilde u_L} = \tilde  m^2_{\tilde d_L}$,
{\it etc}. are guaranteed by the relations (\ref{2.11}) and (\ref{2.12}).
The vacuum expectation values $v_1$ and $v_2$ are given by
\bqa
    v_1 &=& 2{{M_W}\over{M_Z}}\sqrt{M_Z^2-M_W^2}
            {1\over{e\sqrt{1+\tan^2\beta}}}, \cr
    v_2 &=& 2{{M_W}\over{M_Z}}\sqrt{M_Z^2-M_W^2}
            {{\tan\beta}\over{e\sqrt{1+\tan^2\beta}}}. \label{2.16}
\eqa
Note that  the masses of Higgs particles, neutralinos and charginos
are determined  by the mass diagonalization
matrices  introduced in (\ref{2.4}), (\ref{2.6}) and (\ref{2.7}),
respectively.
In particular, the Higgs masses are given at tree level by
\bqa
   M_{H^0,h^0}^2 &=&{1\over 2}\Bigl[ M_{A^0}^2+M_Z^2\pm
             \sqrt{(M_{A^0}^2+M_Z^2)^2-4M_{A^0}^2M_Z^2\cos^22\beta}\Bigr],
             \cr
   M_{H^\pm}^2 &=& M_{A^0}^2+M_W^2. \label{2.17}
\eqa


   According to (\ref{2.17}), the MSSM at Born order predicts
the existence of a CP-even Higgs particle which is lighter than
$Z$.  This possibility having  already been experimentally excluded,
we have to make a minimal modification of the Higgs potential
so that  $M_{H^0}$, $M_{h^0}$ {\it etc}.  can be treated as
free parameters.
For this purpose, inspired by the two-Higgs-doublet model
\cite{gunion,arhrib},
we introduce the effective Higgs potential (see also \cite{semenov,bs})
in the following way:
\bqa
    V^{\rm eff} &=&~ C_1{\bf H}^*_1{\bf H}_1+ C_2{\bf H}^*_2{\bf H}_2 +
           2C_3{\rm Re}({\bf H}_1 {\bf H}_2)
           + 2C_4[{\rm Im}({\bf H}_1 {\bf H}_2)]^2\nonumber \\
       & &+ C_5 ({\bf H}^*_1{\bf H}_1 -{\bf H}^*_2{\bf H}_2)^2
         +C_6\vert{\bf H}^*_1{\bf H}_2\vert^2
         +C_7({\bf H}^*_1{\bf H}_1)^2 + C_8({\bf H}^*_2{\bf H}_2)^2
         \nonumber \\
       & & +4C_9({\bf H}_1^*{\bf H}_1){\rm Re}({\bf H}_1{\bf H}_2)
           +4C_{10}({\bf H}_2^*{\bf H}_2){\rm Re}({\bf H}_1{\bf H}_2)
         \nonumber \\
       & &+C_{11}.  \label{2.18}
\eqa
When the coefficients $C_i$ take special values
\bqa
   C_1 &=&  \tilde m_1^2 +\mu^2
        =M_{A^0}^2\sin^2\beta -{{M_Z^2}\over 2}\cos 2\beta,\label{2.19} \\
   C_2 &=&  \tilde m_2^2 +\mu^2
        =M_{A^0}^2\cos^2\beta +{{M_Z^2}\over 2}\cos 2\beta,\label{2.20} \\
   C_3 &=&  \tilde m_{12}^2
        =-\sin\beta\cos\beta M_{A^0}^2,\label{2.21} \\
   C_5 &=&  \frac{1}{8}(g^2+g^{\prime 2}),  \label{2.22} \\
   C_6 &=& {{g^2}\over 2}, \label{2.23} \\
   C_4 &=& C_7 = C_8 = C_9 = C_{10}=C_{11}= 0, \label{2.24}
\eqa
the above effective Higgs potential reduces to the standard one of the MSSM.
Since $C_{11}$ does not contribute to the Higgs masses nor
to the interactions, the effective Higgs potential depends on ten parameters
$C_i, (i=1\sim 10)$, allowing more freedom than in the case of the original
MSSM Higgs potential which contains three parameters,
$\tilde m_1^2$, $\tilde m_2^2$ and $\tilde m_{12}^2$.
The conditions at the potential minimum relate $C_i$ to $v_j, (j=1,2)$.
Therefore, the effective Higgs potential contains eight free input parameters
\footnote{ Compared with the MSSM, the effective Higgs potential
(\ref{2.18}) contains seven more parameters  and consequently we need seven more
new inputs.}, for which we take
\bq
     M_{A^0},~~~M_{H^0}, ~~M_{h^0},~~~M_{H^\pm},~~~\tan\alpha,~~~
      C_4,~~~C_9,~~~C_{10}.\label{2.25}
\eq
with the convention adopted in {\tt GRACE/SUSY}
\bq
        -{\pi \over 2} \le \alpha \le 0.   \label{2.26}
\eq
In terms of our input parameters, the coefficients $C_i$ can be expressed
as
\bqa
  C_1 &=& -{1\over 4}\bigl[M_{H^0}^2+M_{h^0}^2 + D \cos 2\alpha \bigr]
           - {{v_2}\over{4v_1}} D \sin 2\alpha
           -{{v_2}\over{v_1}}C_3, \label{2.27} \\
  C_2 &=& -{1\over 4}\bigl[M_{H^0}^2+M_{h^0}^2 - D \cos 2\alpha \bigr]
           - {{v_1}\over{4v_2}} D \sin 2\alpha
           -{{v_1}\over{v_2}}C_3, \label{2.28}  \\
  C_3 &=& -\sin\beta\cos\beta M_{A^0}^2 +v_1v_2 C_4 -v_1^2C_9-v_2^2C_{10},
            \label{2.29} \\
  C_5 &=& - {{D\sin 2\alpha}\over{4v_1v_2}}
          -{3\over 2}{{M_{A^0}^2}\over {v_1^2+v_2^2}}
          - {{C_3}\over{v_1v_2}}+{3\over 2}C_4, \label{2.30}\\
  C_6 &=& 2({{M_{H^\pm}^2-M_{A^0}^2}\over{v_1^2+v_2^2}}
         +C_4), \label{2,31} \\
  C_7 &=& -{1\over{v_1^2}}\Bigl[
             C_1 + {{v_2}\over{v_1}}C_3 + (v_1^2-v_2^2)C_5
             +3v_1v_2C_9+{{v_2^3}\over{v_1}}C_{10}\Bigr],
           \label{2.32}\\
  C_8 &=& -{1\over{v_2^2}}\Bigl[
             C_2 + {{v_1}\over{v_2}}C_3 + (v_2^2-v_1^2)C_5
           +{{v_1^3}\over{v_2}}C_9 + 3v_1v_2C_{10}\Bigr],
            \label{2.33}
\eqa
where the vacuum expectation values $v_i$ are given by (\ref{2.16})
\footnote{ (\ref{2.32}) and (\ref{2.33}) are the conditions for the
vanishing of the linear terms in the effective Higgs potential,
thus assuring that $v_1$ and $v_2$ correspond to the minimum of
the potential (\ref{2.18}). } and
\bq
   D = M_{H^0}^2-M_{h^0}^2.       \label{2.34}
\eq

Combining (\ref{2.8}) and (\ref{2.25}), we use the following quantities
as the inputs of {\tt GRACE/SUSY} in the gauge and
Higgs sectors,
\bqa
  &&\alpha_e, \alpha_s, M_W, M_Z, \tan\alpha, \tan\beta, M_{A^0}, M_{H^0},
  M_{h^0}, M_{H^\pm}, \mu, M_1,M_2,M_3, \nonumber \\
  && C_4, C_9, C_{10}.  \label{2.35}
\eqa
In addition to these quantities, all the widths of the unstable particles
must also be specified as input parameters.
The set of the basic input parameters used in the program system are
summarized in Table 1.  The magnitudes of these inputs can be
changed freely by the user. A list of the names of the variables created by
the system in {\tt FORTRAN} codes are given in Appendix B.

\begin{table}
\caption{Input parameters used in {\tt GRACE/SUSY}.}
\begin{tabular}{|l|l|}\hline

     coupling constants  & $\alpha_e$, $\alpha_s$   \\
\hline
     Gauge boson masses  & $M_Z$, $M_W$   \\
\hline
                    & $M_{H^0}$ , $M_{h^0}$, $M_{A^0}$,$M_{H^\pm}$  \\
     Higgs sector   & $\tan\alpha$   \\
                    & $C_4$, $C_9$, $C_{10}$ \\
\hline
     SUSY parameters     & $\mu$, $\tan\beta$, $M_1$, $M_2$, $M_3$  \\
\hline
     fermion masses      & $m_u$, $m_c$, $m_t$  \\
                         & $m_d$, $m_s$, $m_b$  \\
                         & $m_e$, $m_\mu$, $m_\tau$  \\
\hline
     sfermion masses     & $m_{\tilde u_1}$, $m_{\tilde u_2}$,
                             $m_{\tilde c_1}$, $m_{\tilde c_2}$,
                             $m_{\tilde t_1}$, $m_{\tilde t_2}$  \\
                         & $m_{\tilde d_1}$, $m_{\tilde s_1}$,
                             $m_{\tilde b_1}$    \\
                         & $m_{\tilde e_1}$, $m_{\tilde e_2}$,
                             $m_{\tilde \mu_1}$, $m_{\tilde \mu_2}$,
                             $m_{\tilde \tau_1}$, $m_{\tilde \tau_2}$ \\
\hline
     sfermion mixing angles & $\theta_u$, $\theta_c$, $\theta_t$,  \\
                            & $\theta_d$, $\theta_s$, $\theta_b$,  \\
                            & $\theta_e$, $\theta_\mu$, $\theta_\tau$ \\
\hline
     particle widths & $\Gamma_W$, $\Gamma_Z$, ... \\
\hline
\end{tabular}
\end{table}

A possible minimal extension of the MSSM is obtained by setting
\bq
     C_4=C_9=C_{10}=0,
\eq
and using for  $\tan\alpha$ and $M_{H^\pm}^2$ the same form
as given by the tree MSSM,
\bqa
     \tan 2\alpha &=& \tan 2\beta {{M_{A^0}^2+M_Z^2}\over
                                 {M_{A^0}^2-M_Z^2}}, \label{2.37}\\
     M_{H^\pm}^2 &=& M_W^2+M_{A^0}^2,  \label{2.38}
\eqa
respectively.
Concerning $M_{H^0}$ and $M_{h^0}$, one can use
any desired value, such as those evaluated by the one-loop MSSM.

In {\tt GRACE/SUSY}, two modes are
prepared for the choice of Higgs parameters.   In mode 1 the
strict MSSM Higgs potential is used, while in mode 2 the
extended Higgs potential, (\ref{2.18}), is used with the input
parameters given in (\ref{2.25}). 
The mode and  the  input parameters of the
Higgs potential  must be specified in subroutine {\tt sethptal}.
Note that in editing the subroutine, one can also use a different
set of input parameters for  the Higgs potential. See also section
4.4.2.


\subsection{Treatment of the fermion number clashing vertices}
    In models of  SUSY the fermion number may not be conserved at the
interaction vertices when they couple to neutralinos and charginos.
Note that since the fermion number of chargino is not determined by their
interactions, we {\em define\/} the positively charged
charginos as  Dirac fermions.

    In {\tt GRACE/SUSY} we manipulate the fermion
number clashing interactions by introducing the concept
of G-line ({\tt GRACE}-line), F-direction (fermion-direction) and
M-direction (momentum-direction).
The detail of the treatment for the tree case being given in ref.\cite{mk2},
we present here only a brief sketch of the procedure.

Since the flow of a spin one-half particle is not disconnected
nor has any branch, all the external spinors are connected
pairwise.
When such a spinor line connecting two external spinors is
stretched in a straight line and displayed horizontally,
we call it a G-line.
According to the G-line rule to be discussed below,
we compute the G-line
amplitude corresponding to each G-line.
The Feynman amplitude for each graph is then obtained
by multiplying all the so-constructed G-line amplitudes,
the bosonic parts which connect
vertices in different G-lines and the relative
phase of the graph at issue.


\noindent\underbar{G-line rule }\\
Let us assume that the process contains $n_i$ incoming spinors and
$n_f$ outgoing spinors. We assign the particle number 1 to $n_i$
to the $n_i$ incoming spinor particles, and $n_i+1$ to $n_i+n_f$
to the $n_f$ outgoing spinor particles. For a given fermion line
connecting two external spinor particles with particle number $m$
and $n$ ( $m>n$), we define a G-line in such a way that the
particle $m$ comes from the left and the particle $n$ comes from
the right endpoint. This defines the G-line uniquely for each
fermion line belonging to a Feynman graph and there are
$(n_i+n_f)/2$ such G-lines at tree level.

On each segment of the G-line separated by the
interaction vertices, we assign two directions,
an M-direction and an
F-direction \cite{kt}.  For the notational convenience, we also
introduce the notion of an A-direction, which is the direction
in which spinors are ordered in the
Feynman amplitude when a fermion comes in and goes out after
some interactions, namely, the direction indicated by the
left-oriented arrow.

\noindent{\bf M-direction} \\
An M-direction is assigned in a natural way.
For external incoming particles, their M-direction is the
direction towards the vertex, while for the external outgoing
particles it is the direction from the vertex.
For the internal fermions, the M-direction is always taken
in the A-direction (irrespective of whether the internal
propagator is that of fermion or anti-fermion when seen in the
A-direction).

\noindent{\bf F-direction} \\
For Dirac particles we assign the F-direction in a natural way;
for an external fermion the F-direction is the same as its
M-direction, while for an external anti-fermion it is opposite to
its M-direction. For the internal Dirac particle it is taken to be
equal to the A-direction for fermions (when seen in the
A-direction) and to the opposite direction for anti-fermions.
Concerning  Majorana particles which are self-charge-conjugate,
how to assign their F-direction is a matter of convention. In our
convention, we define the F-direction of  Majorana particles by
their M-direction.
This means  that the F-direction
of the internal Majorana particle  coincides with its
A-direction and only the following type of the
Majorana propagator appears in the G-line amplitude;
\bq
   <\Psi \bar\Psi> = \bullet \leftarrow \leftarrow \bullet
                   = {i\over{\gslash p -m}},
\eq
where the momentum $p$ runs from right to left.
Clashing of the F-direction occurs only at vertices, and the
charge conjugation matrix appears only at the fermion number
clashing vertices.

The above assignments of the F-direction and
the M-direction uniquely identify the external spinors as shown
in Table 2.
\begin{table}
\caption{Spinor assignment at the left and right endpoint of the G-lines.}
\begin{tabular}{|l|l|l|l|l|}\hline
  F-direction  & $\longleftarrow\bullet$ & $\longrightarrow\bullet$
               & $\bullet\longleftarrow$ & $\bullet\longrightarrow$ \\
  M-direction  &  &  &  &  \\ \hline
  $\longleftarrow$ & $\bar u(p)$ & $v(p)^T$ & $u(p)$ & $\bar v(p)^T$ \\
  $\longrightarrow$& $\bar v(p)$ & $u(p)^T$ & $v(p)$ & $\bar u(p)^T$ \\
\hline
\end{tabular}
\end{table}

The general rule of flipping one of the F-direction at the
vertex is summarized as follows
\bqa
  \longleftarrow\bullet --- &=& -C(\longrightarrow\bullet ---), \nonumber \\
  \longrightarrow\bullet --- &=& -C^{-1}(\longleftarrow\bullet ---),
       \nonumber \\
  ---\bullet\longleftarrow &=& (---\bullet\longrightarrow)C^{-1}, \nonumber \\
  ---\bullet\longrightarrow &=& (---\bullet\longleftarrow)C, \nonumber
\eqa
where $---$ is a fermion line whose F-direction can be arbitrary.
The charge conjugation matrix $C$ satisfies the following properties
\bq
    C^\dagger= C^{-1},~~~C=-C^T,~~~C[1,\gamma_5,\gamma_\mu,\gamma_\mu\gamma_5]^T
    C^{-1} = [1,\gamma_5,-\gamma_\mu,\gamma_\mu\gamma_5].
\eq
It is also important to note that we can choose the relative phase of the
spinor wave functions such that
\bq
    C\bar v(k,\lambda)^T = u(k,\lambda),~~~~
    C\bar u(k,\lambda)^T = v(k,\lambda). \label{2.41}
\eq

The rule of reversing the particle order at a vertex is given as
follows;
\bqa
     B\longrightarrow\bullet\longrightarrow A &=&
     -(A \longleftarrow\bullet\longleftarrow B)^T, \nonumber \\
     B\longrightarrow\bullet\longleftarrow A &=&
     -(A \longrightarrow\bullet\longleftarrow B)^T, \nonumber\\
     B\longleftarrow\bullet\longrightarrow A &=&
     -(A \longleftarrow\bullet\longrightarrow B)^T. \nonumber
\eqa

Applying the above rules to the lagrangian of
fermion number conserving Dirac fermions interactions
(including chargino),
\bq
  {\cal L}_{DD} = \overline\Psi(D_\alpha)\Gamma_{D_\alpha,D_\beta}
                    \Psi(D_\beta) + h.c., \label{2.42}
\eq
one obtains the G-line rule for the Dirac particles
(neglecting the overall factor $i$)
\bqa
   D_\alpha \longleftarrow\bullet\longleftarrow D_\beta
    & \qquad\qquad & \Gamma_{D_\alpha,D_\beta}, \\
   D_\beta \longrightarrow\bullet\longrightarrow D_\alpha
    & \qquad\qquad & -\Gamma^T_{D_\alpha,D_\beta}.
\eqa
Concerning the vertices including Majorana particles, corresponding
to the lagrangian
\bqa
  {\cal L}_{MD} &=& \overline\Psi(M_\alpha)\Gamma_{M_\alpha,D_\beta}
                    \Psi(D_\beta)
                  + \overline\Psi(D_\beta)\Gamma_{D_\beta,M_\alpha}
                    \Psi(M_\alpha), \label{2.45}\\
  {\cal L}_{MM}&=&\sum_{\alpha\ge\beta}
            \overline\Psi(M_\alpha)\Gamma_{M_\alpha,M_\beta}
                    \Psi(M_\beta),  \label{2.46}
\eqa
where $\Gamma_{D_\alpha, M_\beta} \equiv
\gamma_0\Gamma^\dagger_{M_\beta,D_\alpha}\gamma_0$ and $D$ stands for
a Dirac particle which can be a chargino, we obtain the following G-line rule,
\bqa
   M_\alpha \longleftarrow\bullet\longleftarrow D_\beta
    & \qquad\qquad & \Gamma_{M_\alpha,D_\beta}, \\
   D_\alpha \longleftarrow\bullet\longleftarrow M_\beta
    & \qquad\qquad & \Gamma_{D_\alpha,M_\beta}, \\
   M_\alpha \longleftarrow\bullet\longleftarrow M_\beta
    & \qquad\qquad & \Gamma_{M_\alpha,M_\beta}, ~~(\alpha\ge\beta) \\
   M_\beta \longleftarrow\bullet\longrightarrow D_\alpha
    & \qquad\qquad & C\Gamma^T_{D_\alpha,M_\beta}, \label{2.50}\\
   D_\beta \longrightarrow\bullet\longleftarrow M_\alpha
    & \qquad\qquad & -\Gamma^T_{D_\beta,M_\alpha}C^{-1}. \label{2.51}
\eqa
Because the F-direction of Majorana particles is
fixed to be equal to their A-direction, other vertices
with Majorana particle having F-directions opposite to the
A-direction do not appear in the G-lines.
We also need the G-line rule for the fermion number
violating chargino interactions
with a conventional Dirac particle, whose lagrangian is given by
\bq
   {\cal L}_{CD} = \overline\Psi^c(C)\Gamma_{C,D}\Psi(D)
                  +\overline\Psi(D)\Gamma_{D,C}\Psi^c(C),
\eq
where $ \Gamma_{D,C} = \gamma_0\Gamma_{C,D}^\dagger\gamma_0$.
The rules are
\bqa
  C \longrightarrow \bullet \longleftarrow D 
   & \qquad\qquad &  -C^{-1}\Gamma_{C,D},
     \\
  D \longleftarrow \bullet \longrightarrow C 
   & \qquad\qquad &  \Gamma_{D,C}C.
    \label{2.54}
\eqa
Table 2 and eq.(\ref{2.42})-(\ref{2.54}) are what we need to compute the
G-line amplitudes.

As Denner et al. \cite{denner} have done, it is possible
to formulate the fermion number
clashing G-line amplitude without using the charge conjugation matrix $C$
if one observes that the propagator with its F-direction opposite to
the A-direction is given by
\bqa
   \bullet\rightarrow\rightarrow\bullet
   &=& -S^T(-k) = -C^{-1}S(k)C,  \\
   \leftarrow k~~~ & &   \nonumber
\eqa
and that the right-hand side of (\ref{2.50}) and (\ref{2.51}) is
expressed in terms of the charge conjugated vertex function
\bq
    \Gamma^c_{A,B} \equiv C\Gamma_{B,A}C^{-1},
\eq
as
\bqa
    M_\beta \longleftarrow\bullet\longrightarrow D_\alpha
      & \qquad\qquad &  \Gamma^c_{M_\beta,D_\alpha}C, \\
    D_\beta \longrightarrow\bullet\longleftarrow M_\alpha
      & \qquad\qquad &  -C^{-1}\Gamma^c_{D_\beta,M_\alpha},
\eqa
In ref.\cite{mk2} the equivalence of these two approaches has been
demonstrated.

\section{Comments}
     Several comments are in order.

We have checked and debugged the possible coding errors in {\tt
GRACE/SUSY} by several methods.  First, the interaction vertices
coded in the system, which are based on the expression presented
in \cite{mk1}, are compared with  the result  computed by the
algebraic manipulation program {\tt REDUCE}. As another check, we
have used the gauge invariance of the processes which
include  photon, $W^{\pm}$, $Z$ and gluon. In this check one has
to make the widths of the heavy gauge bosons vanish.  By changing
the value of gauge parameters, we have confirmed at one phase space
point that the amplitudes of  all the possible 582,102 processes
with up to 6 external legs remain unchanged within the error of 1
in $10^{15}$ or less in double-precision, and 1 in $10^{30}$ or
less in quadruple-precision calculation. Also further checks have
been made by Boudjema and B\'elanger of LAPTH \cite{lappcpp}, who
compared the {\tt GRACE/SUSY} results of several SUSY processes
with those of {\tt CompHEP} \cite{comphep}, the automatic
computation program for the high energy processes developed by a
group of Moscow State University.

Using {\tt GRACE/SUSY} we have computed a number of
SUSY processes and some of the results have been published.
As this system  allows us to deal with any SUSY particle production process at various
types of
high energy colliders, there are analyses for
$e^+e^-$
\cite{susy23,Lafage:1999mv,Belanger:1998ip,Belanger:1999rq,Belanger:2000ky},
$ep$
\cite{Kurihara:1998kv,Matsushita:1998ku,Kitamura:2001mu,Kon:2000hc},
$e\gamma$, $\gamma\gamma$
\cite{Watanabe:1998mn,Kon:2001hm,Badelek:2001xb,Abe:2001gc}
and
$pp$ colliders \cite{Abdullin:1999zp}.
Note, moreover, that some works have been done with a version of
{\tt GRACE/SUSY} extended to the $R$-parity violating SUSY models
\cite{Matsushita:1998ku,Kitamura:2001mu,Kon:2000hc,Kon:2001hm,%
Badelek:2001xb,Abe:2001gc,rbphys}.

    {\tt GRACE/SUSY} can treat as many final-state particles as
possible, provided that one prepares by oneself the kinematics routine
necessary for the phase space integration and
the computer has an enough capacity to store
the gigantic output files.  The system is, however, equipped with
its own kinematics library which has all the kinematics routines
necessary for the analysis of up to 4-body final-state
(see Appendix C).

    Users can modify the program for their own interest
but at their own risk.
In particular, if one would like to use the other set
of input parameters for SUSY lagrangian,
for example, the trilinear coupling constants $A_f$
in place of $\theta_f$, one should prepare a
program in which the  parameter values
are converted from the user's set to that of
{\tt GRACE/SUSY} given in Table 1,
and call the program before the amplitude calculation starts.
Note that in the system it is $m_fA_f$ but not $A_f$
itself that has the variable name (see Appendix B).
Note also that in the sfermion sector, the unmixed physical
state is obtained by setting either $\theta_f=0$ or
$\theta_f= {\pi/2}$.  In the former case, however,
the identification corresponds to $\tilde f_1 =\tilde f_L$
and $\tilde f_2=\tilde f_R$, while in the latter
$\tilde f_1 =\tilde f_R$
and $\tilde f_2=\tilde f_L$.  Since $m_{\tilde f_1}\le m_{\tilde f_2}$
is assumed in the system, we have to set
$\theta_f=0$ $(\theta_f={\pi/2})$ when one considers
the case in which there is no mixing in the sfermion sector and
the left-handed sfermion is lighter (heavier) than the right-handed one.

The program {\tt GRACE/SUSY} is designed strictly within the
framework of the tree level.  To be more realistic, however, one
can arrange the value of the input parameters so that they are
subject to the renormalization group equations (RGE). A program
{\tt MUSE} developed by Lafage \cite{Belanger:1998ip,lafage} can
compute the SUSY particle masses as solutions of the coupled RGE.
One can combine such a program with {\tt GRACE/SUSY} to invoke
improved values for the input parameters. See also
ref.\cite{FeynHiggs}.

   It is also possible to extend {\tt GRACE/SUSY} so as to include
new particles or
new interactions such as the $R$-parity violating interaction
(see, for example, \cite{Kon:2000hc}).

The algorithm of G-line works without any change
for loop diagrams.
Hence the extension of {\tt GRACE/SUSY} for one-loop
process is possible once one had appropriate loop
evaluation routines.

In its current version of {\tt GRACE/SUSY} the quark-mixing matrix is taken
to be the unit matrix. Neutrinos are massless and left-handed.
One may introduce the quark
mixing by modifying the model files in an appropriate way.


\section{How to use {\tt GRACE/SUSY}}

   In this section we show how to use the program {\tt GRACE/SUSY}
by adopting a SUSY  process
$e^+e^- \to \gamma\tilde\chi_1^+\tilde\chi_1^-$ as an example.
This process has
28 Feynman graphs at tree level, among which 4 graphs contain
even number of fermion number violating vertices.

\subsection{ Definition of physical process }
    First the physical process to be computed must be defined,
which is given in the file "{\tt in.prc}".
The content of the file {\tt in.prc} for the process
$e^+e^- \to \gamma\tilde\chi_1^+\tilde\chi_1^-$ is as follows;
\par
\begin{quote}
\begin{verbatim}
Model="mssm.mdl";
Process;
ELWK=3;
QCD=0;
  Initial={electron,positron};
  Final  ={photon,chargino1,anti-chargino1};
Kinem="2302";
Pend;
\end{verbatim}
\end{quote}
\par

\subsubsection{\bf Model file }
In the first line the model file is defined.
Two model files are prepared, one is {\tt mssm.mdl} for the
minimal SUSY standard model and the other is {\tt ms.mdl} for
the standard model.  The latter will be useful when one wants to
separate the SUSY contributions from those of non-SUSY
in such processes where only the SM particles appear
in the external legs.
\par
\subsubsection{\bf Order of perturbation }
The orders of perturbation are to be defined; {\tt ELWK} for
the electroweak interaction and {\tt QCD} for the QCD.
\par
\subsubsection{\bf Initial and final states }
In the lines for the initial and final states, the particle names
of the process are to be given for each state according to the
list given in Appendix A. One should be careful about the ordering
of the particles (in particular, of massless gauge bosons) in each
state. The program reads this input list and numbers all the
particles from 1 to $n$ for $n$ external legs according to the
order in which they appear in these two lines. These assigned
numbers are in turn used to specify the integral variables in the
phase space integration by the kinematic library. For example, for
the 3-body final state including a single photon, the photon must
appear first in the final state specification as our example
shows. Otherwise the best choice of the integral variables will
not be taken. See the next subsection and Appendix C for detail.
In our example of $e^+e^- \to \gamma\tilde\chi_1^+\tilde\chi_1^-$
the input file {\tt in.prc} defines the particle numbers {\tt
1}$\sim${\tt 5} as $e^-$ ($p_1$), $e^+$ ($p_2$), $\gamma$ ($p_3$),
$\chi_1^+$ ($p_4$) and $\chi_1^-$ ($p_5$).
\par


\subsubsection{\bf Kinematics code number}
 {\tt GRACE/SUSY} has a built-in kinematics library which
contains  a set of kinematics routines for the
phase space integration.
In order to specify the kinematics routine to be used,  one has to choose
an appropriate code number, {\tt 2302} in our example.
A list of code numbers is given in Appendix C.
\par
\subsection{ Graph generation}
The graph generation procedure starts by typing the command {\tt grc}:
\par
\begin{quote}
\begin{verbatim}
grace% grc
\end{verbatim}
\end{quote}
\par
\noindent
All Feynman graphs for the process are generated and their information
is saved in the file {\tt out.grf}.

\subsection{ Drawing graphs }
A Feynman graph drawer {\tt gracefig} is initiated by the command:
\begin{quote}
\begin{verbatim}
grace% gracefig
\end{verbatim}
\end{quote}
\par
\begin{figure}
\begin{center}
 \epsfbox{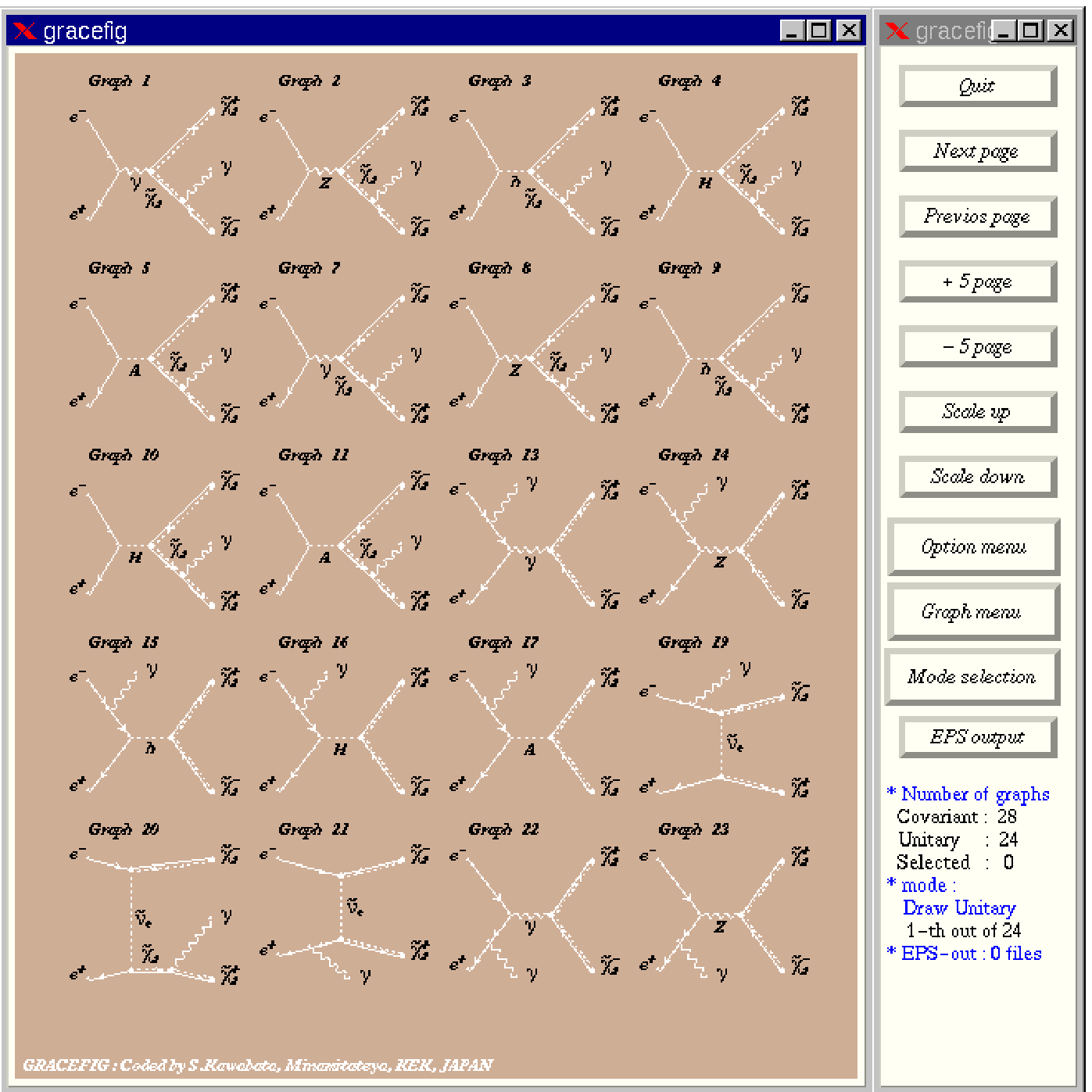}
 \caption{
  Feynman graphs for $e^+e^-$ $\to$
  $\gamma\tilde{\chi_1^+}\tilde{\chi_1^-}$
  in the MSSM.
  }
 \label{fig:feynman}
\end{center}
\end{figure}

\noindent
The procedure {\tt gracefig} reads the graph information from the
file {\tt out.grf} and analyzes the structure of each graph,
and then creates two windows.  The output of our example process
is shown in Fig.{\ref{fig:feynman}}.

\subsubsection{ Drawing window of {\tt gracefig}}
On the drawing window the first $m \times n$ Feynman graphs
 are drawn, in the unitary gauge by default, where the number of rows $m$ and
 the number of columns $n$ are determined
automatically by the size of screen and the total number of
graphs. When the size of the window is changed, the number of
columns $n$ is conserved while the number of rows $m$ is adjusted
so as to fit  the height of the window.
\par

\subsubsection{ Control window of {\tt gracefig}}
In the control window several buttons are set with which one can
select through the pull-down menus several options which are
listed in Table 3.

Particularly one can draw all the Feynman graphs in the covariant gauge by
choosing an appropriate button on the menu.

\par
\subsubsection{ To draw selected graphs }
There are two modes, ``{\em Drawing Mode\/}'' and ``{\em Select Mode\/}''.
When {\tt gracefig} is invoked, all Feynman graphs in the unitary
gauge are displayed in the {\em drawing mode\/}.
If one wants to see or print only some specific graphs, one
can select them in the {\em select mode\/} by clicking their
pictures in the drawing window.
The background color of the selected graphs are darkened in this mode.
After this selection only the selected graphs can be
 displayed in the {\em drawing mode\/}
by choosing ``{\em Selected graphs\/}'' in the graph menu.

\begin{table}
\caption{Buttons and sub-menus of \texttt{gracefig}}
\begin{tabular}{|l|p{9cm}|}
\hline
 Button~/~Sub-menu        &   Function \cr
\hline
{\bf Quit}                &  Exit from {\tt gracefig}.         \cr
{\bf Next page}           &  Show next page.                    \cr
{\bf Previous page }      &  Show previous page.                \cr
{\bf \verb/+/5 page }     &  Show 5-th page ahead.              \cr
{\bf \verb/-/5 page }     &  Show 5-th page back.               \cr
{\bf Scale up }           &  Make the size of graph larger.
       The number of graphs in a row: $n  \rightarrow  n-1$.   \cr
{\bf Scale down }         &  Make the size of graph smaller.
       The number of graphs in a row: $n  \rightarrow  n+1$.   \cr
{\bf Option menu}         &  Pull down menu.                    \cr
{\em ~~/~Graph number}    &  On/Off display of graph number.    \cr
{\em ~~/~Particle number} &  On/Off display of particle number. \cr
{\em ~~/~Vertex number}   &  On/Off display of vertex number.   \cr
{\em ~~/~Particle name}   &  On/Off display of particle name.   \cr
{\em ~~/~Line number}     &  On/Off display of internal line number. \cr
{\bf Graph menu}          &  Pull down menu.                    \cr
{\em ~~/~Covariant Gauge} &  Show graphs in covariant gauge.    \cr
{\em ~~/~Unitary Gauge}   &  Show graphs in unitary gauge (Default). \cr
{\em ~~/~Selected Graphs} &  Show only the selected graphs.     \cr
{\bf Mode Selection}      &  Pull down menu.                    \cr
{\em ~~/~Drawing Mode}    &  Default mode to draw the graphs.   \cr
{\em ~~/~Select Mode}     &  To select a graph, click on a graph you
                                                            choose. \cr
{\bf EPS output}          &  Output the current page on a file
                             {\tt grcfig*.eps} in eps format.      \cr
\hline
\end{tabular}
\end{table}

\subsubsection{ EPS output and insertion of a picture in a text}
When ``EPS output'' button is clicked, the EPS file for the current page
in the drawing window is created.
The name of EPS file is ``{\tt grcfig{\it nn}.eps}'', where {\it nn} is
the sequential number starting from 0, and one EPS file corresponds
to one page of the graphs.

\subsection{ Source code generation }
After the graph generation,  by typing the command
\par
\begin{quote}
\begin{verbatim}
grace% grcfort
\end{verbatim}
\end{quote}
\par
\noindent
{\tt FORTRAN} code and {\tt Makefile} are generated, which are
necessary to perform gauge independence check, the numerical
integration and the event generation.

\begin{figure}
\setlength{\unitlength}{1mm}
\begin{picture}(150,170)

\put(34,160)  {\framebox(22,8)[c]{\hbox{\large \bf BASES}}}
\put(33,159)  {\framebox(24,10)[c]{}}
\put(78,160)  {\framebox(24,8)[c]{\hbox{\large \bf SPRING}}}
\put(77,159)  {\framebox(26,10)[c]{}}

\put(0,153)  {\framebox(20,8)[c]{\hbox{\small \tt mainbs}}}
\put(120,153){\framebox(20,8)[c]{\hbox{\small \tt mainsp}}}
  \put(10,153) {\line(0,-1){149}}
  \put(130,153){\line(0,-1){149}}

\put(15,60)  {\dashbox(20,8)[c] {\hbox{\small \tt BASES}}}
\put(15,40)  {\dashbox(20,8)[c]{\hbox{\small \tt BSINFO}}}
\put(15,20)  {\dashbox(20,8)[c]{\hbox{\small \tt BHPLOT}}}
\put(15,0)   {\dashbox(20,8)[c]{\hbox{\small \tt BSWRIT}}}
  \put(10, 64) {\line(1,0){5}}
  \put(10, 44) {\line(1,0){5}}
  \put(10, 24) {\line(1,0){5}}
  \put(10, 4)  {\line(1,0){5}}

\put(105,110){\dashbox(20,8)[c]{\hbox{\small \tt BSREAD}}}
\put(105,60) {\dashbox(20,8)[c] {\hbox{\small \tt SPRING}}}
\put(105,20) {\dashbox(20,8)[c]{\hbox{\small \tt SPINFO}}}
\put(105,0)  {\dashbox(20,8)[c]{\hbox{\small \tt SHPLOT}}}
  \put(130,114){\line(-1,0){5}}
  \put(130,64) {\line(-1,0){5}}
  \put(130,24) {\line(-1,0){5}}
  \put(130,4)  {\line(-1,0){5}}

\put(60,140){\dashbox(20,8)[c] {\hbox{\small \tt bsinit}}}
  \put(10, 144){\line(1,0){50}}
  \put(130,144){\line(-1,0){50}}

  \put(10, 134){\line(1,0){50}}
  \put(130,84){\line(-1,0){35}}
  \put(95,84){\line(0,1){50}}
  \put(95,134){\line(-1,0){15}}
\put(60,130){\framebox(20,8)[c]{\hbox{\small \tt userin}}}
\put(60,120){\framebox(20,8)[c]{\hbox{\small \tt setmdl}}}
\put(60,110){\framebox(20,8)[c]{\hbox{\small \tt setcpl}}}
\put(60,100){\framebox(20,8)[c]{\hbox{\small \tt sethptal}}}
\put(60,90) {\framebox(20,8)[c]{\hbox{\small \tt gfinit}}}
\put(60,80) {\dashbox(20,8)[c]{\hbox{\small \tt kinit}}}
  \put(55, 132){\line(0,-1){48}}
  \put(55, 132){\line(1,0){5}}
  \put(55, 124){\line(1,0){5}}
  \put(55, 114){\line(1,0){5}}
  \put(55, 94) {\line(1,0){5}}
  \put(55, 84) {\line(1,0){5}}
  \put(85, 114){\line(0,-1){10}}
  \put(85, 114){\line(-1,0){5}}
  \put(85, 104){\line(-1,0){5}}

  \put(35, 64){\line(1,0){25}}
  \put(105,64){\line(-1,0){25}}
\put(60,60) {\framebox(20,8)[c]{\hbox{\small \tt func}}}
  \put(60,70) {\dashbox(20,8)[c]{\hbox{\small \tt kfill}}}
  \put(60, 66){\line(-1,0){5}}
  \put(55, 66){\line(0,1){8}}
  \put(55, 74){\line(1,0){5}}

  \put(60,50) {\dashbox(20,8)[c]{\hbox{\small \tt kinem}}}
  \put(60,40) {\framebox(20,8)[c]{\hbox{\small \tt ampsum}}}
  \put(60,30) {\framebox(20,8)[c]{\hbox{\small \tt amptbl}}}
  \put(60,20) {\framebox(20,8)[c]{\hbox{\small \tt amps}}}
  \put(60,10) {\framebox(20,8)[c]{\hbox{\small \tt ampord}}}
  \put(60, 62){\line(-1,0){5}}
  \put(55, 62){\line(0,-1){28}}
  \put(55, 54){\line(1,0){5}}
  \put(55, 44){\line(1,0){5}}
  \put(55, 34){\line(1,0){5}}
  \put(70, 30){\line(0,-1){2}}
  \put(70, 20){\line(0,-1){2}}

  \put(80, 24){\line(1,0){10}}
  \put(90, 24){\line(0,-1){20}}
  \put(90, 4){\line(-1,0){5}}
\put(55,0) {\dashbox(30,8)[c]{}}
\put(57,4) {{\footnotesize Interface and }}
\put(57,1) {{\footnotesize \texttt{CHANEL} Library}}
\end{picture}
\caption{Relation among the program components.}
\end{figure}
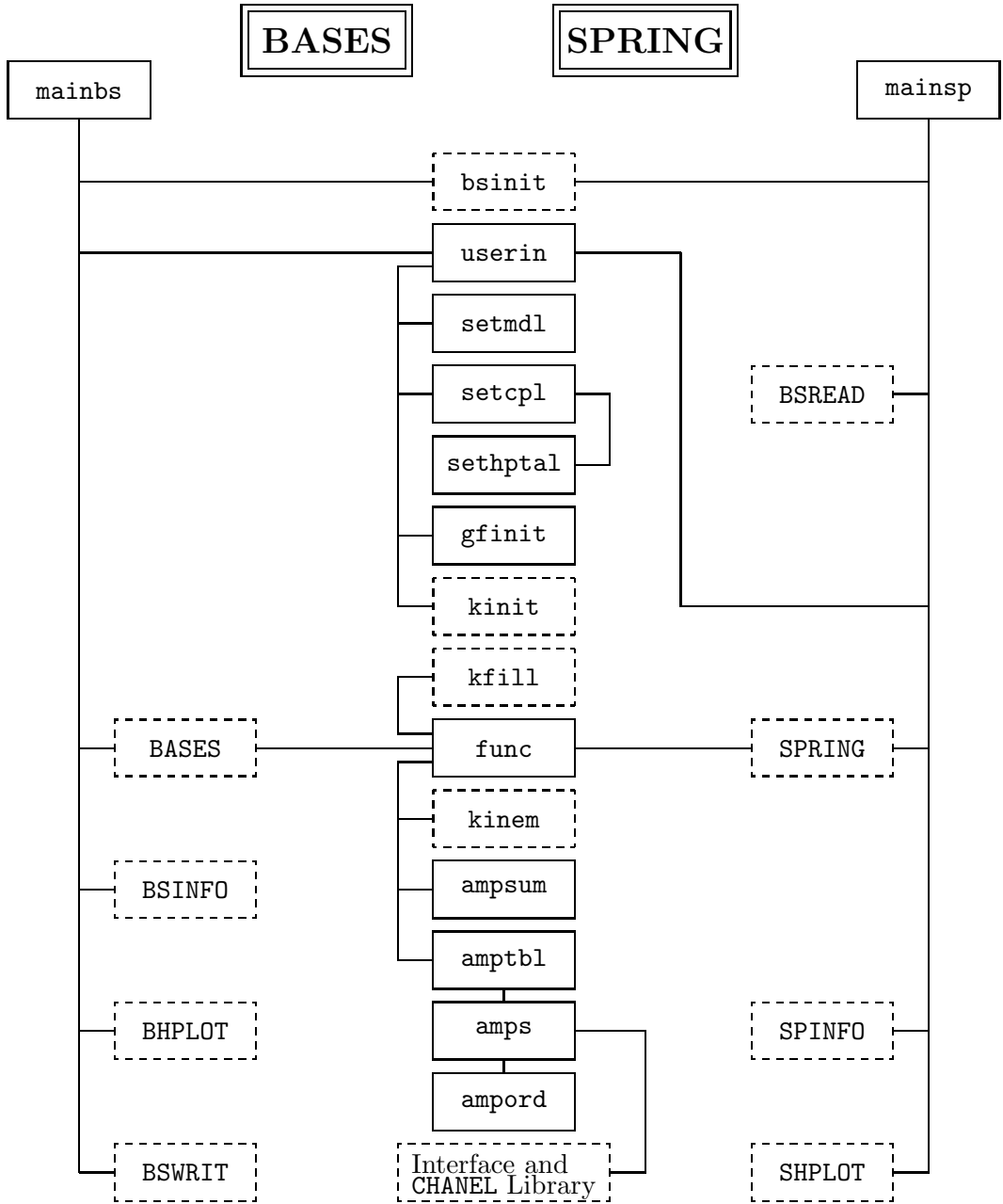

\subsubsection{Generated source code}
Three kinds of program components are created by the system. They are
used in  the amplitude calculation,
in the phase space integration by {\tt BASES} and in the event
generation by {\tt SPRING}, respectively.


The interrelation among the subprograms generated by {\tt GRACE/SUSY} is
depicted in Fig.2, where those subprograms in the {\em solid box\/} are
 automatically
generated by {\tt GRACE/SUSY}, while those in the {\em dashed box\/} are
already contained in other program packages
{\tt BASES/SPRING}, kinematics library, interface program library to
{\tt CHANEL} and program package {\tt CHANEL}.
This figure omits a few minor components.
The components vertically located in the middle of
the figure are commonly used in {\tt BASES} and {\tt SPRING}.
The program specifications of {\tt BASES/SPRING}, the
interface and {\tt CHANEL} are
described in references \cite{grace,kon1,tkkjk,mtc1}.
A brief description of generated subprograms is given below:

\begin{itemize}
\item[(1)] Main programs. \\

\noindent
\begin{tabular}{p{1.5cm}p{2.5cm}p{8.5cm}}
\hline
{\tt mainbs} &{\it (  main       )}& is the main program for the integration.\\
{\tt mainsp} &{\it (  main       )}& is the main program for the event generation.\\
{\tt gauge}  &{\it (  main       )}& is the main program for the check
of gauge independence. \\
\hline
\end{tabular}
\par
\item[(2)] A set of program components for the integration by {\tt BASES} and
for the event generation by {\tt SPRING}.
\par
\noindent
\begin{tabular}{p{1.5cm}p{2.5cm}p{8.5cm}}
\hline
{\tt userin} &{\it (  subroutine )}& controls initialization. \\
{\tt setmdl} &{\it (  subroutine )}& defines MSSM parameters, masses and 
                                     decay widths of particles.\\
{\tt sethptal} &{\it ( subroutine )}& defines the parameters of the extended 
                                      Higgs potential.\\
{\tt setcpl} &{\it (  subroutine )}& defines coupling constants.\\
{\tt gfinit} &{\it (  subroutine )}& initializes particle polarization and 
                                     selects Feynman graphs (amplitudes).\\
{\tt kinit}  &{\it (  subroutine )}& initializes integration parameters,
histograms, kinematics and user's parameters.\\
\end{tabular}
\begin{tabular}{p{1.5cm}p{2.5cm}p{8.5cm}}
\hline
{\tt func}   &{\it (  function   )}& calculates the numerical
values of differential cross section.\\
{\tt kinem}  &{\it (  subroutine )}& derives particle four momenta from the
                                     integration variables.\\
{\tt kfill}  &{\it (  subroutine )}& fills values in histograms and 
                                     scatter-plots.\\
\end{tabular}
\begin{tabular}{p{1.5cm}p{2.5cm}p{8.5cm}}
\hline
{\tt amptbl} &{\it (  subroutine )}& calls {\tt amnnnn} to calculate 
                                     amplitudes.\\
{\tt ampsum} &{\it (  subroutine )}& sums the square of the matrix elements \\
             &                    & over the helicity states.  \\
{\tt amps} &{\it (  subroutine )}& calculates amplitudes.\\
{\tt ampord} &{\it (  subroutine )}& arranges amplitudes.\\
\end{tabular}
\begin{tabular}{p{1.5cm}p{2.5cm}p{8.5cm}}
\hline
{\tt incl1.h}  &{\it (  include file )}& defines the common variables
                  for masses, amplitude tables {\it etc}.\\
{\tt incl2.h}  &{\it (  include file )}& defines the work space for
{\tt amptbl}.\\
{\tt inclk.h}  &{\it (  include file )}& transfers values of masses and
a few constants.\\
\end{tabular}
\begin{tabular}{p{1.5cm}p{2.5cm}p{8.5cm}}
\hline
{\tt userout} &{\it (  subroutine )}& prints the amplitude summary table.
Used only by {\tt gauge}.\\
\hline
\end{tabular}
\par

\end{itemize}
\par

The subprograms {\tt userin} and {\tt func} are used both in the numerical
integration by {\tt BASES} and in  the event generation by {\tt SPRING}.
In {\tt userin} the subroutine {\tt kinit} is
called to initialize the kinematics routine  and other parameters.
The subroutines {\tt setmdl}, {\tt setcpl} and {\tt gfinit} are also called for
the initialization
of physical parameters and amplitude calculation.

The function subprogram {\tt func} is used for calculating the numerical value
of differential cross section, where the subroutine {\tt kinem} is called for
calculating four-momentum vectors of external particles, and
{\tt amptbl} and
{\tt ampsum} are called for the amplitude calculation.
The detailed specifications of these program components are
described in refs. \cite{grace,mtc1}.

\subsubsection{ Setting physical parameters for numerical calculation}
Prior to running the program, one has to set the values
of the input physical parameters
by editing appropriate {\tt FORTRAN} source codes.
The parameters and corresponding subroutines to be edited
are categorized as follows.
\par
\begin{description}
\item[\underline{kinematical parameters}]  \qquad {\tt kinit}, {\tt kfill}
\item[\underline{MSSM parameters}] \qquad {\tt setmdl}, {\tt sethptal}
\item[\underline{polarization and graph selection}] \qquad {\tt gfinit}
\end{description}

\par
\begin{itemize}
\item[(1)] {\bf Initialization routine {\tt kinit}}
        \par
        The values of the following parameters can be changed
        by editing {\tt kinit}.
        \begin{itemize}
        \item[(a)] Modification of physical parameters
            \begin{itemize}
            \item Center of mass energy in GeV.
              $\sqrt{s} = 1000$GeV is a default setting, which corresponds to
                a description {\tt w = 1000.d0} in {\tt kinit}. \par
                \item Various kinematical cuts {\it etc}.
            \item Option flags in built-in kinematics routines (\cite{kinemlib}).
            \end{itemize}
        \item[(b)] Integration parameters which control the action of
         {\tt BASES} can be altered in {\tt kinit}.
         The following parameters can be changed from their default values
         according to the required
         integration accuracy:
\par
\begin{tabular}{lp{10cm}}
{\tt ncall}                   & The number of sampling points per iteration\\
{\tt acc1}                    & The expected accuracy for the grid optimization step\\
{\tt itmx1}                   & The maximum iteration number for the grid optimization\\
{\tt acc2}                    & The expected accuracy for the integration step\\
{\tt itmx2}                   & The maximum iteration number for the integration step\\
\end{tabular}
\par
         However, the following parameters are usually not to be changed:
\par
\begin{tabular}{lp{10cm}}
{\tt ndim}                    & The dimensions of integral \\
{\tt nwild}                   & The number of wild variables\\
{\tt xl({\it i}),xu({\it i})} & The lower and upper bounds of integration
                                variable {\tt x({\it i})}\\
{\tt ig({\it i})}             & The grid optimization flag for {\it i}-th variable\\
\end{tabular}
\par

        \item[(c)] Initialization of histograms and scatter-plots\\
        If the user wants histograms for some physical variables,
        the initialization routines {\tt xhinit} and {\tt dhinit} should be
        called for each histogram and scatter-plot, respectively.
        \par
        The system automatically generates source code to call these
        routines so as
        to construct histograms with respect to each integration variable,
        {\tt x(i), i=1} $\sim$ {\tt ndim} and the energies and angles of
        final particles.
        If the user does not need these default histograms, one may
        delete the corresponding {\tt FORTRAN} statements in {\tt kinit}.
        When the user needs more histograms for some distributions,
        one should add lines to call {\tt xhinit} or {\tt dhinit}.
        These routines are to be written in the following format
        \begin{verbatim}
 call xhinit( id#,
.             lower_limit, upper_limit, # of bins,
.             'Title of this histogram' )
        \end{verbatim}
        and
        \begin{verbatim}
 call dhinit( id#,
.             x-lower_limit, x-upper_limit, # of x bins,
.             y-lower_limit, y-upper_limit, # of y bins,
.             'Title of this scatter plot' )
        \end{verbatim}
            where $1\le \verb/id#/ \le 99$.
        \end{itemize}
\item[(2)] {\bf Filling histograms and scatter-plots in {\tt kfill}} \par
        In the generated {\tt kfill} by {\tt GRACE/SUSY}, histograms for all
        integration variables and scatter-plots for
        all combinations of them
        are filled  by calling {\tt xhfill} and {\tt dhfill},
        respectively.
        If one changes the initialization of histograms
        and scatter-plots in {\tt kinit},
        their filling parts should be changed accordingly.
        These routines are to be called as
        \begin{verbatim}
         call xhfill( id#, v, func )
         call dhfill( id#, vx, vy, func )
        \end{verbatim}
\item[(3)]  {\bf Subprogram {\tt setmdl}} \par
       The MSSM parameters, masses and widths of
         particles are defined in {\tt setmdl} subroutine.
         The variable names of all the parameters are listed in Appendix B.
         As an example we show how to set the MSSM parameters in the case
         ($\mu$, $M_2$, $\tan\beta$, $m_{\tilde{\nu_e}}$) =
         ($-300$GeV, $200$GeV, $10$, $150$GeV).
        \begin{verbatim}
          tanbe   =     10.d0
          xmu     =   -300.d0
          xm2     =    200.d0
          amsn(1) =    150.d0
        \end{verbatim}

\item[(4)]  {\bf Subprogram {\tt sethptal}}\par
       In this subprogram one has to specify the Higgs potential
       to be used in the
       calculation by assigning a value 1 or 2 to the variable {\tt ihiggs}.
       See section 2.4 for detail.
        \begin{itemize}
        \item   {\tt ihiggs = 1}:   mode 1; MSSM Higgs potential
        \item   {\tt ihiggs = 2}:   mode 2; general extended Higgs potential
                                            (default)
        \end{itemize}
        The default value is {\tt ihiggs=2}, in which case
        one has to give values of the additional seven input parameters,
        for example,
         \begin{verbatim}
          amh1  = 115.d0
          amh2  = 300.d0
          amhc  = 350.d0
          al    = -0.37d0
          c4    = 0.d0
          c9    = 0.d0
          c10   = 0.d0
        \end{verbatim}

\item[(5)]  {\bf Subprogram {\tt gfinit}}
       \begin{itemize}
        \item[(a)] Selection of graphs. \par
         The contribution from a
         graph is calculated when its flag is set to be {\tt jselg = 1 }
         in {\tt gfinit}.
         For example, if one wants to know the contributions only
         from the sneutrino exchange
         (graphs 19, 20, 21 and 28 in Fig.\ref{fig:feynman}), one should
         write
        \begin{verbatim}
      do 10 n1 = 1, ngraph
        jselg(n1) = 0
   10 continue
        jselg(19) = 1
        jselg(20) = 1
        jselg(21) = 1
        jselg(28) = 1
        \end{verbatim}
         The selection of graphs should respect gauge invariance, otherwise
the cross section is physically not sensible.
         It should be noted that this specification of graphs is independent
         from the graph
         selection in the Feynman graph drawer {\tt gracefig}.
    \item[(b)] Set of polarization. \par
        As the helicity amplitude technique is used in {\tt GRACE/SUSY}, the
        beam polarization can be specified.
        The polarization sum is performed by setting the flags
        {\tt jhs(i)} and {\tt jhe(i)} for particle $i$ having spin $s_i$,
        where $0\le {\tt jhs(i)} \le {\tt jhe(i)} \le 2s_i$.
       In default the unpolarized cross section is assumed, which corresponds to
        setting the flags for the initial $e^-$ (particle 1) and
        $e^+$ (particle 2) as
        \begin{verbatim}
*     1: initial electron mass=amlp(1)
         ...........
      jhs(1) = 0
      jhe(1) = 1
         ............
*     2: initial positron mass=amlp(1)
         ...........
      jhs(2) = 0
      jhe(2) = 1
         ...........
        \end{verbatim}
       When we consider the initial $e^-$ beam polarization, for example,
        they should be changed to
        \begin{verbatim}
      jhs(1) = 0
      jhe(1) = 0
        \end{verbatim}
        for helicity=-1 and
        \begin{verbatim}
      jhs(1) = 1
      jhe(1) = 1
        \end{verbatim}
        for helicity=+1.
       \end{itemize}

      Similarly, one can choose polarization vectors of external gauge bosons
      by putting {\tt jhs(i), jhe(i) = 0, 1}(transverse),
      {\tt 2}(longitudinal: only
      for massive gauge bosons).  In the default the linearly polarized
      states are realized as combinations of the transverse polarizations.
      When one wants to take the circular polarization for a photon,
      an option switch {\tt jcpol} should be changed from {\tt 0} to {\tt 1}.

    \item[(c)] Gauge parameters. \par
      Without a specification the unitary gauge is used in the
      integration over the phase space.  However, one can
      select covariant gauge (\(R_\xi\)-gauge)
      with the gauge parameters having arbitrary values.

      The selection of gauge is controlled by integer variables
      \texttt{igauwb}, \texttt{igauzb}, \texttt{igauab} and \texttt{igaugl}
      which correspond to gauge bosons \(W^\pm, Z, \gamma\) and
      gluon, respectively.
      When these parameters are set to be \texttt{0},
      unitary gauge is selected.
      Otherwise they should be \texttt{igauwb = 1}, \texttt{igauzb = 2},
      \texttt{igauab = 3} and \texttt{igaugl = 4}.
      The value of each gauge parameter should be given by
      \texttt{agauge(igau}\textsl{xx}\texttt{)},
      where \textsl{xx} is one of \texttt{wb}, \texttt{zb}, \texttt{ab}
      or \texttt{gl}.

      In the file \texttt{gfinit.f} one will find the following lines which
      select the unitary gauge.
\begin{verbatim}
* Gauge parametes (default is unitary gauge)
      igauab = 0
      igauwb = 0
      igauzb = 0
      igaugl = 0

      agauge(0) = 1.0d20
\end{verbatim}
      In a general covariant gauge these lines should be changed
      as follows;
\begin{verbatim}
* Gauge parametes (covariant gauge)
      igauwb = 1
      igauzb = 2
      igauab = 3
      igaugl = 4

      agauge(igauwb) = ...
      agauge(igauzb) = ...
      agauge(igauab) = ...
      agauge(igaugl) = ...
\end{verbatim}
\end{itemize}

\subsection{ Compilation and gauge independence check }

As described in a previous subsection, {\tt Makefile} is also generated by the
command {\tt grcfort}. The command {\tt make} creates the executable files
{\tt gauge}, {\tt integ} and {\tt spring}.

Before calculating the cross section a preliminary test of the
program source should be done. In {\tt GRACE/SUSY} system 
gauge independence test is prepared for this purpose. The numerical
values of the differential cross section at one sample point in
the phase space are calculated both in the covariant and unitary
gauge, and the consistency between these two values is examined.
By executing {\tt gauge}, the contributions to the cross section
at one phase space point
in unitary gauge ({\tt ans1}) and in covariant gauge ({\tt ans2}), and their
difference ({\tt ans1/ans2-1}) are shown as the following example
demonstrates.
\par
\begin{quote}
{\footnotesize \begin{verbatim}
    . . . . .
   ans1   =  0.1255918104665936E+01
    . . . . .
          ...............
   ans2   =  0.1255918104665936E+01
    . . . . .
          ...............
   ans1/ans2 - 1 =  2.22044605E-16
    . . .. . .
\end{verbatim}}
\end{quote}
\par
The relative difference should be around
$10^{-16}$ in the double-precision computation and around $10^{-30}$
in the quadruple-precision.

It should be noted that even though this gauge independence check
shows consistency between the two gauges, it does not necessarily
guarantee a complete gauge independence, since it tests only at a
specific point in the phase space. It is recommended to repeat a
similar check at several points in  phase space by changing  the
variable {\tt x(i)}  in {\tt gauge.f} from their default value of
{\tt 0.45} to other value lying in the range {\tt 0 < x(i) < 1}.

\subsection{Integration}

After
a successful test of gauge independence,
one can proceed to the numerical integration by {\tt BASES}, whose details
can be found in ref.\cite{grace}.
By the {\tt make} command one has already created
the executable files {\tt integ} for the integration and {\tt spring} for
event generation.

For integration the command {\tt integ} is used.

\begin{quote}
{\small \begin{verbatim}
grace% integ
\end{verbatim}}
\end{quote}

The integration parameters
and the result of the integration at each iteration
are shown on the screen,  and at the same time they are
automatically stored in the file {\tt bases.result} together with the
histogram outputs. If one uses a built-in kinematics routine,
the cross section is given in unit of {\em pico-barn\/}(pb).

The numerical integration is performed by two successive steps,
the first is the grid optimization step and the second the
integration step. In the former the gridding of the phase space in
hypercubes is optimized for the integrand, while in the latter an
accurate estimate of the integral is evaluated with frozen grids
determined in the first step.

Before termination of the integration procedure, {\tt BASES}
creates a binary file {\tt bases.data} and writes the probability
information on it, which is used for the event generation.

The system will issue {\tt WARNING} messages when the convergence
is not well established. It is always recommended to look at the
integration result carefully, particularly over the convergence
behavior both for the grid optimization and integration step. When
the convergence accuracy fluctuates largely from one iteration to
another, and/or, when  in some cases it suddenly jumps up to a
value far from the preceding iterations, the resultant estimate of
integral is not reliable. There are two possible origins of such a
behavior; one is due to too small number of sampling points and
the other, more serious, is due to an unsuitable choice of the
integration variables for the integrand. In the latter case 
one should use a better set of the integral variables.
\par
\subsection{Event generation}
The event generation starts with the command
\begin{quote}
{\small \begin{verbatim}
grace% spring
\end{verbatim}}
\end{quote}
\par

\noindent
Then {\tt SPRING} reads the probability information from
the binary file {\tt bases.data} and asks the number of events
with the following prompt:

\begin{quote}
{\small \begin{verbatim}
Number of events ?
\end{verbatim}}
\end{quote}

\par\noindent
The user must type the number of events to be generated.
\par
If the sample point does not meet the generation criteria, such a point is discarded
as a "failed generation".
The event generation will run until a given number of events are
generated or the number of failed generations
exceeds its given maximum {\tt mxtry}, whose default value is 50.
\par
In order to estimate the computing time for the event generation, one can
refer to  the expected generation time given in the information of
{\tt BASES} output, where the generation efficiency of 70\% is assumed.
\par
\subsubsection{ Analysis of events }
In the event generation four-momentum vectors of the final particles
are automatically stored in {\tt vec(i,n)} in {\tt mainsp},
where {\tt n} stands for the particle number and ${\tt i =1\sim 4}$
correspond to the three-momentum ({\tt 1}, {\tt 2}, {\tt 3})
and the energy ({\tt 4}) of the particle.

To create histograms of some variables of the generated events,
one has to initialize them before the generation loop and
fill them after each call of {\tt spring}.
This kind of histograms is called as an ``{\em additional histogram\/}''
in contrast to the ``{\em original histogram\/}'' which is filled
both in the integration and event generation steps.
In the original histograms
one can see how the generated events reproduce those
distributions produced by the integration, while in the additional
histogram only the frequency distribution can be seen as
in the usual histogram.
An example for saving the four-momentum vectors and creating
additional histograms is given below.
\par
\begin{quote}
\begin{verbatim}

*     Open the output file for the events
      . . . . . .

*     Initialize the additional histograms by
        call xhinit( id, ... ).

      do 100 nevnt = 1, mxevnt

         call spring( func, mxtry )

*    ------------------------------------------------
*        analyze the event and
*        fill the additional histograms by
          call xhfill( id, .... ).
*    ------------------------------------------------

         do 90 k = 1 , nextrn
            write(6,*) (vec(j,k),j=1,4)
   90    continue
  100 continue
\end{verbatim}
\end{quote}
\par
\subsubsection{ Output from {\tt SPRING}}
The output from {\tt SPRING} is written on the file named {\tt
spring.result}, which consists of the general information,
original and additional histograms, scatter-plots, and the
number-of-trials distribution. In the last distribution one can
see the generation efficiency. It is recommended to check if this
distribution has a sharp peak in the first bin. If this is not the
case, the integration should be carried out again with more
sampling  points or with more iterations.


\begin{ack}
   We would like to thank F.~Boudjema and G.~B\'elanger
for checking parts of the program and reading the manuscript.
We would like to thank also D.~Perret-Gallix for his interest
in this work and for encouragement.
This work was partly supported by Japan Society for Promotion of
Science under the Grant-in-Aid for Scientific Research B ( No.11440083 ).
\end{ack}



\newpage
\appendix
\section{List of particles in MSSM}

  The particles appearing in MSSM and their variable names are listed in two
tables below.  These variable names are to be used in specifying the process
in {\tt in.prc}.


\begin{tabular}{|l|l|r|}\hline
   particle & variable name \\
\hline
   photon       &  {\tt photon} \\
   $W^+ (W^-)$  &  {\tt W-plus}  ({\tt W-minus})  \\
   $Z$          &  {\tt Z}     \\
   gluon        &  {\tt gluon} \\
\hline
   $\nu_e (\overline{\nu_e})$  &   {\tt nu-e} ({\tt nu-e-bar})\\
   $e^- (e^+)$    &   {\tt electron} ({\tt positron}) \\
   $\nu_\mu (\overline{\nu_\mu})$ & {\tt nu-mu} ({\tt nu-mu-bar})\\
   $\mu^- (\mu^+)$     &  {\tt muon} ({\tt anti-muon})\\
   $\nu_\tau (\overline{\nu_\tau})$& {\tt nu-tau} ({\tt nu-tau-bar})\\
   $\tau^- (\tau^+)$    &  {\tt tau} ({\tt anti-tau}) \\
\hline
   $u (\bar{u})$       &   {\tt u} ({\tt u-bar}) \\
   $d (\bar{d})$       &   {\tt d} ({\tt d-bar})  \\
   $c (\bar{c})$       &   {\tt c} ({\tt c-bar})\\
   $s (\bar{s})$       &   {\tt s} ({\tt s-bar})\\
   $t (\bar{t})$       &   {\tt t} ({\tt t-bar})\\
   $b (\bar{b})$       &   {\tt b} ({\tt b-bar})\\
\hline
   $h^0$   & {\tt Higgs1}  \\
   $H^0$   & {\tt Higgs2}  \\
   $A^0$   & {\tt Higgs3}  \\
   $H^+ (H^-)$  &  {\tt Higgs-plus} ({\tt Higgs-minus})  \\
\hline
\end{tabular}

\newpage


\begin{tabular}{|l|l|r|}\hline
   particle & variable name \\
\hline
   $\tilde{\chi^+_1} (\tilde{\chi^-_1})$ & {\tt chargino1} ({\tt anti-chargino1})\\
   $\tilde{\chi^+_2} (\tilde{\chi^-_2})$ & {\tt chargino2} ({\tt anti-chargino2})\\
\hline
   $\tilde\chi^0_1$ & {\tt neutralino1} \\
   $\tilde\chi^0_2$ & {\tt neutralino2} \\
   $\tilde\chi^0_3$ & {\tt neutralino3} \\
   $\tilde\chi^0_4$ & {\tt neutralino4} \\
\hline
   $\tilde{\nu_e} (\overline{\tilde{\nu_e}})$ & {\tt snu-e} ({\tt anti-snu-e}) \\
   $\tilde{\nu_\mu} (\overline{\tilde{\nu_\mu}})$ & {\tt snu-mu} ({\tt anti-snu-mu}) \\
   $\tilde{\nu_\tau} (\overline{\tilde{\nu_\tau}})$ & {\tt snu-tau} ({\tt anti-snu-tau}) \\
   $\tilde{e^-_1} (\tilde{e^+_1})$ & {\tt selectron1} ({\tt anti-selectron1}) \\
   $\tilde{e^-_2} (\tilde{e^+_2})$ & {\tt selectron2} ({\tt anti-selectron2}) \\
   $\tilde{\mu^-_1} (\tilde{\mu^+_1})$   & {\tt smuon1} ({\tt anti-smuon1}) \\
   $\tilde{\mu^-_2} (\tilde{\mu^+_2})$   & {\tt smuon2} ({\tt anti-smuon2}) \\
   $\tilde{\tau^-_1} (\tilde{\tau^+_1})$ & {\tt stau1} ({\tt anti-stau1}) \\
   $\tilde{\tau^-_2} (\tilde{\tau^+_2})$ & {\tt stau2} ({\tt anti-stau2}) \\
\hline
   $\tilde{u_1} (\overline{\tilde{u_1}})$  &   {\tt su1} ({\tt anti-su1}) \\
   $\tilde{u_2} (\overline{\tilde{u_2}})$  &   {\tt su2} ({\tt anti-su2}) \\
   $\tilde{d_1} (\overline{\tilde{d_1}})$  &   {\tt sd1} ({\tt anti-sd1}) \\
   $\tilde{d_2} (\overline{\tilde{d_2}})$  &   {\tt sd2} ({\tt anti-sd2}) \\
   $\tilde{c_1} (\overline{\tilde{c_1}})$  &   {\tt sc1} ({\tt anti-sc1}) \\
   $\tilde{c_2} (\overline{\tilde{c_2}})$  &   {\tt sc2} ({\tt anti-sc2}) \\
   $\tilde{s_1} (\overline{\tilde{s_1}})$  &   {\tt ss1} ({\tt anti-ss1}) \\
   $\tilde{s_2} (\overline{\tilde{s_2}})$  &   {\tt ss2} ({\tt anti-ss2}) \\
   $\tilde{t_1} (\overline{\tilde{t_1}})$  &   {\tt st1} ({\tt anti-st1}) \\
   $\tilde{t_2} (\overline{\tilde{t_2}})$  &   {\tt st2} ({\tt anti-st2}) \\
   $\tilde{b_1} (\overline{\tilde{b_1}})$  &   {\tt sb1} ({\tt anti-sb1}) \\
   $\tilde{b_2} (\overline{\tilde{b_2}})$  &   {\tt sb2} ({\tt anti-sb2}) \\
\hline
\end{tabular}


\section{Name List of {\tt GRACE/SUSY} Parameters}

The variable names used in {\tt GRACE/SUSY} are summarized.
All mass variables. widths and the dimensional coupling constant $\mu$
must be given in unit of GeV.  Those variables whose magnitude can
be specified by the user in {\tt setmdl.f} and {\tt sethptal.f} are
indicated by an
asterisk "*" at the end of the row. {\bf
The value of other variables should not be changed by the user.
Such a modification may cause internal inconsistencies.  }

\noindent \underbar{gauge bosons}

\noindent
\begin{tabular}{|l|l|l|c|}\hline
  category         & notation      & variable name in {\tt GRACE/SUSY}
       & \\ \hline
  gauge boson masses  & $M_Z$, $M_W$  & {\tt amz}, {\tt amw} & * \\
  gauge boson widths  & $\Gamma_Z$, $\Gamma_W$
                      & {\tt agz}, {\tt agw}  & *\\
  coupling            &  $\alpha_e$, $\alpha_s$ & {\tt alpha}, {\tt alphas} &*\\
                      & $e$,  $g_s$ & {\tt ge}, {\tt gs} &  \\
                      & $g$, $g^\prime$ & {\tt gg}, {\tt gz} & \\
  mixing angle        & $\cos\theta_W$, $\sin\theta_W$, $\tan\theta_W$
                      & {\tt cw}, {\tt sw}, {\tt tw} &  \\
\hline
\end{tabular}

\noindent\underbar{Ordinary fermions}

\noindent
\begin{tabular}{|l|l|l|c|}\hline
 fermion mass    & $m_u$, $m_c$, $m_t$
                 & {\tt amuq(1)}, {\tt amuq(2)}, {\tt amuq(3)} & *\\
                 & $m_d$, $m_s$, $m_b$
                 & {\tt amdq(1)}, {\tt amdq(2)}, {\tt amdq(3)}  & *\\
                 & $m_e$, $m_\mu$, $m_\tau$
                 & {\tt amlp(1)}, {\tt amlp(2)}, {\tt amlp(3)} & * \\
\hline
 fermion widths  & $\Gamma_{u}$, $\Gamma_{c}$, $\Gamma_{t}$
                 & {\tt aguq(1)}, {\tt aguq(2)}, {\tt aguq(3)} & * \\
                 & $\Gamma_{d}$, $\Gamma_{s}$, $\Gamma_{b}$
                 & {\tt agdq(1)}, {\tt agdq(2)}, {\tt agdq(3)} & * \\
                 & $\Gamma_{e}$, $\Gamma_{\mu}$, $\Gamma_{\tau}$
                 & {\tt aglp(1)} , {\tt aglp(2)}, {\tt aglp(3)} & * \\
\hline
\end{tabular}
\\
Of course one should set {\tt aguq(1) = agdq(1) = aglp(1) = 0}.


\noindent\underbar{Higgs bosons}

\noindent
\begin{tabular}{|l|l|l|c|}\hline
   SUSY parameters  & $\mu$  & {\tt xmu} & * \\
\hline
   Higgs masses  & $M_{A^0}$     & {\tt amh3}  & * \\
                 & $M_h$, $M_H$, $M_{H^\pm}$  &
                 {\tt amh1}, {\tt amh2}, {\tt amhc} & * \\
\hline
  Higgs widths  & $\Gamma_h$, $\Gamma_H$, $\Gamma_A$, $\Gamma_{H^\pm}$
                & {\tt agh1}, {\tt agh2}, {\tt agh3}, {\tt aghc} & * \\
\hline
  mixing angle  & $\alpha$, $\cos\alpha$, $\sin\alpha$
                & {\tt al}, {\tt ca}, {\tt sa}  & * \\
                & $\cos\beta$, $\sin\beta$,$\tan\beta$, $\cot\beta$
                & {\tt cb}, {\tt sb}, {\tt tanbe}, {\tt cotbe} & *  \\
\hline
\end{tabular}

\noindent\underbar{Chargino}

\noindent
\begin{tabular}{|l|l|l|c|}\hline
  SUSY parameters   & $\mu$, $M_2$ & {\tt xmu}, {\tt xm2} & *\\
\hline
  chargino masses   & $m_{\tilde \chi^\pm_1}$, $m_{\tilde \chi^\pm_2}$,
                    & {\tt amsw(1)}, {\tt amsw(2)} &  \\
\hline
  chargino widths   & $\Gamma_{\tilde \chi^\pm_1}$,
                      $\Gamma_{\tilde \chi^\pm_2}$,
                    & {\tt agsw(1)}, {\tt agsw(2)} & * \\
\hline
  chargino mixing angles &$\sin\phi_L$, $\cos\phi_L$, $\epsilon_L$
                    & {\tt sphl}, {\tt cphl}, {\tt epsl} & \\
                    & $\sin\phi_R$, $\cos\phi_R$
                    & {\tt sphr}, {\tt cphr} &  \\
\hline
\end{tabular}

\noindent\underbar{Neutralino}

\noindent
\begin{tabular}{|l|l|l|c|}\hline
  SUSY parameters    & $\mu$, $M_1$, $M_2$ &
                      {\tt xmu}, {\tt xm1}, {\tt xm2} & * \\
\hline
  neutralino masses  & $m_{\tilde \chi^0_i}$ & {\tt amnl(i)} & \\
\hline
  neutralino widths  & $\Gamma_{\tilde \chi^0_i}$ & {\tt agnl(i)} & * \\
\hline
  neutralino mixing matrix & $({\cal O}_N)_{ij}$, $\eta_i$  &
                            {\tt ogmn(i,j)}, {\tt etan(i)}(complex) & \\
\hline
\end{tabular}

\noindent\underbar{Gluino}

\noindent
\begin{tabular}{|l|l|l|c|} \hline
  gluino mass  & $m_{\tilde g} = M_3$     &
                 {\tt amgl} $=$ {\tt xm3} & * \\
  gluino width & $\Gamma_{\tilde g}$ & {\tt aggl} & * \\
\hline
\end{tabular}

\noindent\underbar{Sfermion}

\noindent
\begin{tabular}{|l|l|l|c|}\hline
  sfermion masses  & $m_{\tilde u_i}$, $m_{\tilde c_i}$, $m_{\tilde t_i}$
                   & {\tt amsu(1,i)}, {\tt amsu(2,i)}, {\tt amsu(3,i)}& *  \\
                   & $m_{\tilde d_i}$, $m_{\tilde s_i}$,
                             $m_{\tilde b_i}$
                   & {\tt amsd(1,i)}, {\tt amsd(2,i)}, {\tt amsd(3,i)}& * \\
                   & $m_{\tilde \nu_e}$, $m_{\tilde \nu_\mu}$,
                          $m_{\tilde \nu_\tau}$
                   & {\tt amsn(1)}, {\tt amsn(2)}, {\tt amsn(3)}& *  \\
                   & $m_{\tilde e_i}$, $m_{\tilde \mu_i}$,
                             $m_{\tilde \tau_i}$
                   & {\tt amsl(1,i)}, {\tt amsl(2,i)}, {\tt amsl(3,i)}& * \\
\hline
  sfermion widths  & $\Gamma_{\tilde u_i}$, $\Gamma_{\tilde c_i}$,
                        $\Gamma_{\tilde t_i}$
                   & {\tt agsu(1,i)}, {\tt agsu(2,i)}, {\tt agsu(3,i)}& * \\
                   & $\Gamma_{\tilde d_i}$, $\Gamma_{\tilde s_i}$,
                             $\Gamma_{\tilde b_i}$
                   & {\tt agsd(1,i)}, {\tt agsd(2,i)}, {\tt agsd(3,i)}&* \\
                   & $\Gamma_{\tilde \nu_e}$, $\Gamma_{\tilde \nu_\mu}$,
                          $\Gamma_{\tilde \nu_\tau}$
                   & {\tt agsn(1)}, {\tt agsn(2)}, {\tt agsn(3)}&* \\
                   & $\Gamma_{\tilde e_i}$, $\Gamma_{\tilde \mu_i}$,
                             $\Gamma_{\tilde \tau_i}$
                   & {\tt agsl(1,i)}, {\tt agsl(2,i)}, {\tt agsl(3,i)}& * \\
\hline
  SUSY-breaking  & $m_uA_u$, $m_cA_c$,  $m_tA_t$
                 & {\tt xauq(1)}, {\tt xauq(2)}, {\tt xauq(3)} & \\
  parameters     & $m_dA_d$, $m_sA_s$,  $m_bA_b$
                 & {\tt xadq(1)}, {\tt xadq(2)}, {\tt xadq(3)} & \\
                 & $m_eA_e$, $m_\mu A_\mu$, $m_\tau A_\tau$
                 & {\tt xalp(1)}, {\tt xalp(2)}, {\tt xalp(3)} & \\
\hline
  sfermion mixing  &$\sin\theta_{\tilde u}$, $\sin\theta_{\tilde c}$,
  angles            $\sin\theta_{\tilde t}$
                   & {\tt shuq(1)}, {\tt shuq(2)}, {\tt shuq(3)} &*\\
                   &$\cos\theta_{\tilde u}$, $\cos\theta_{\tilde c}$,
                    $\cos\theta_{\tilde t}$
                   & {\tt chuq(1)}, {\tt chuq(2)}, {\tt chuq(3)} &*\\
                   &$\sin\theta_{\tilde d}$, $\sin\theta_{\tilde s}$,
                    $\sin\theta_{\tilde b}$
                   & {\tt shdq(1)}, {\tt shdq(2)}, {\tt shdq(3)} &*\\
                   &$\cos\theta_{\tilde d}$, $\cos\theta_{\tilde s}$,
                    $\cos\theta_{\tilde b}$
                   & {\tt chdq(1)}, {\tt chdq(2)}, {\tt chdq(3)} &*\\
                   &$\sin\theta_{\tilde e}$,$\sin\theta_{\tilde \mu}$,
                    $\sin\theta_{\tilde \tau}$
                   & {\tt shlp(1)}, {\tt shlp(2)}, {\tt shlp(3)} &*\\
                   &$\cos\theta_{\tilde e}$,$\cos\theta_{\tilde \mu}$,
                    $\cos\theta_{\tilde \tau}$
                   & {\tt chlp(1)}, {\tt chlp(2)}, {\tt chlp(3)} &*\\
\hline
\end{tabular}


\section{Kinematics Library}

\subsection{Cross section formula }
The volume element for each of final particles is given by
\begin{equation}
\int{{\rm d}^3p\over2E(2\pi)^3},
\end{equation}
where ${\bf p}$ is the three momentum of the particle and $E$ is its
energy.

In a collision of two particles labeled by 1 and 2, the flux is given by
\begin{equation}
           \hbox{flux}=v_{rel}2E_12E_2,\quad
           v_{rel}=\left\vert{{\bf p}_1\over E_1}
                           -{{\bf p}_2\over E_2} \right\vert,
\end{equation}
with $v_{rel}$ being the relative velocity of particles 1 and 2.
The total cross section is then written as
\begin{equation}
\sigma = {1\over\hbox{flux}}\prod_j\int{{\rm d}^3p_j\over2E_j(2\pi)^3}
(2\pi)^4\delta^4(p_1+p_2-\sum_jp_j)\sum_{h_f}\sum_{h_i}|T_{if}|^2,
\end{equation}
where $T_{if}$ is the helicity amplitude for the process $i\to f$.
Initial helicities are averaged and the final ones are summed.
This formula is used for the cross section in the {\tt GRACE/SUSY} system.
The cross section is given in the unit of
{\sl pico-barn}(pb) in the kinematics library.

For the decay of a particle with momentum $p$, the total decay rate
is given by
\begin{equation}
\Gamma = {1\over2E}\prod_j\int{{\rm d}^3p_j\over2E_j(2\pi)^3}
(2\pi)^4\delta^4(p-\sum_jp_j)\sum_{h_f}\sum_{h_i} |T_{if}|^2.
\end{equation}

The independent integration variables expressed in terms of  momenta
and angles must be mapped onto those of the integration package
{\tt BASES} which are normalized as $ 0\le {\tt x(i)} \le 1$, where {\tt i}
runs from 1 to {\tt ndim},  the dimension of the integration,

{\bf One should remember that the statistical factor for the identical
particles is not generated automatically and should be supplied
by hand.} This is because this factor depends on the observables one is
interested in. For the total cross section one should divide the result
obtained by {\tt BASES} by the factorial of the number of identical
particles.

\subsection{Kinematics library}
{\tt GRACE/SUSY} assumes the following two types of  processes:
\[
p_1+p_2 \rightarrow p_3+p_4+\cdots
   \quad\quad (\hbox{2-body scattering}),
\]
\[
p_1\rightarrow p_2+p_3+p_4+\cdots
   \quad\quad (\hbox{decay of a particle}).
\]
It should be noted that the particle assignment follows the order
in which the  particles are originally defined in {\tt in.prc}.

The Lorentz invariant phase space element for a final \(n\)-body
(\(A\rightarrow 1+2+3+\cdots+n\)) is defined by
\bqa
d\tilde\Gamma_n&=&(2\pi)^4\delta^4\left(\sum_{in}p-\sum_{out}^n p\right)
   \prod_{out}^n \frac{d^3p_j}{(2\pi)^3 2E_j}
  \equiv \frac{1}{(2\pi)^{3n-4}} d\Gamma_n, \\
d\Gamma_n&=&\delta^4\left(\sum_{in}p-\sum_{out}^n p\right)
   \prod_{out}^n \frac{d^3p_j}{2E_j}.
\eqa
The following chain rule is useful(\(0<k<n-1\)):
\bqa
   &~& d\Gamma_n(A\rightarrow 1+2+3+\cdots)    \nonumber\\
   &~& \qquad\qquad\qquad = \;
   d\Gamma_{k+1}(A\rightarrow 1+\cdots+k+q) \frac{dQ^2}{2\pi} \nonumber \\
   &~& \qquad\qquad\qquad\quad
     \times d\Gamma_{n-k}(q\rightarrow (k+1)+(k+2)+\cdots),
\eqa
where \(q^2=Q^2\).

When \(p_a\) is in the center-of-mass system, {\it i.e.},
\({\bf p}_a={\bf 0}\), and if \(p_a=p_b+p_c\) we introduce the 2-body
phase space
\bqa
d\Gamma_{\rm CM}(a;bc)\equiv d\Gamma_2
  =\delta^4(p_a-p_b-p_c)\frac{d^3p_b}{2E_b}\frac{d^3p_c}{2E_c},
\eqa
and write it by angular variables:
\bqa
 d\Gamma_{\rm CM}(a;bc)=\frac{\beta(a;bc)}{8}d\Omega_{\rm CM}(a;bc)
 =\frac{\beta(a;bc)}{8}d\cos\theta_{b,(bc)} d\phi_{b,(bc)}.
\eqa
Here $\beta$ is given by
{\footnotesize
\bqa
\beta(a;bc)&=&\frac{2P}{E_a}  \\
       &=&\frac{1}{E_a}
\sqrt{(E_a+m_b+m_c)(E_a-m_b-m_c)(E_a+m_b-m_c)(E_a-m_b+m_c)}. \nonumber
\eqa
}
The subscript \((bc)\) of the angles indicates that
\(\theta_b, \phi_b\) are defined in the center-of-mass frame.
Angles in the laboratory frame have no subscript.

Kinematics routines prepared in the library are summarized in
Table~\ref{kinematics-table}.
\begin{center}
  \begin{tabular}{|c||l|}
  \hline
  code number & \hspace*{5mm}contents \\
  \hline
  \hline
%
%
  {\tt 1201}  & 1-body \(\rightarrow\) 2 body decay \\
  \hline
%
%
  {\tt 1301}  & 1-body \(\rightarrow\) 3 body decay \\
  &Sequential decay \(1\rightarrow 2+(3+4)\rightarrow 2+3+4 \)
can be treated.\\
  \hline
  \hline
%
%
  {\tt 2201}  & 2-body \(\rightarrow\) 2 body in CM frame \\
  & \(t\)- and \(u\)-channel singularities can be treated. \\

  \hline
%
%
  {\tt 2301}  & 2-body \(\rightarrow\) 3 body in CM frame ,\\
  &Sequential decay type \(1+2\rightarrow 3+(4+5)\rightarrow 3+4+5 \).\\
  &  Resonance on particles 4 and 5 can be treated.\\

  {\tt 2302}  & 2-body \(\rightarrow\) 3 body in CM frame ,\\
  & Radiative processes \(1+2\rightarrow 3(\gamma)+4+5\),\\
  & both initial and final radiation can be treated.\\

  {\tt 2303}  & 2-body \(\rightarrow\) 3 body in CM frame ,\\
  &Double-radiative processes \(1+2\rightarrow 3(\gamma)+4(\gamma)+5\)\\

  {\tt 2304}  & 2-body \(\rightarrow\) 3 body in CM frame ,\\
  &Three photon processes \(1+2\rightarrow
                             3(\gamma)+4(\gamma)+5(\gamma)\)\\
  \hline
%
%
  {\tt 2401}  & 2-body \(\rightarrow\) 4 body in CM frame,
  a pair of sequential \\
  & decay type \(1+2\rightarrow (3+4)+(5+6)\rightarrow 3+4+5+6 \)\\
  &  \(t\)-channel singularity can be treated. \\

  {\tt 2402}  & 2-body \(\rightarrow\) 4 body in CM frame,\\
  &`fusion' type \(1+2\rightarrow (3+A)+(4+B); A+B \rightarrow 5+6 \)\\
  \hline
  \end{tabular}
\label{kinematics-table}

Table~\ref{kinematics-table}~{\footnotesize Kinematics library
prepared in {\tt GRACE/SUSY} system.}
\end{center}

\end{document}